\documentclass[11pt]{article}
\usepackage{jheppubmodnew}
\pdfoutput=1
\usepackage{psfrag}
\usepackage{array}
\usepackage{amssymb}
\usepackage{amsmath}
\usepackage{amsthm}
\usepackage{mathtools}
\usepackage[normalem]{ulem}
\usepackage{graphicx}
\usepackage{comment}
\usepackage{cite}
\usepackage{epsfig}
\usepackage{tikz}
\usetikzlibrary{arrows,plotmarks,positioning,calc}
\usepackage{mathrsfs}
\usepackage{bbm}
\usepackage[colorlinks=true]{hyperref}
\usepackage{ragged2e}
\usepackage{tikz}
    \usetikzlibrary{decorations.markings}
    \usetikzlibrary{cd} % commutative diagram package
    
\numberwithin{equation}{section} %section numbers in equations
\usepackage{cite}

\newcommand{\alset}{{\cal A}}
\newcommand{\altc}{{\cal A}_{\text{light}}}
\newcommand{\alqft}{{\cal A}_{\text{QFT}}}
\newcommand{\honedim}{{L^2(\mathbb{R})}}
\newcommand{\albarqft}{\overline{\cal A}_{\text{QFT}}}
\newcommand{\obsen}{E}
\newcommand{\obsenfluct}{\delta E}

\newcommand{\obsst}{\psi}
\newcommand{\smobsst}{\Psi}

\newcommand{\fullst}{f,\mathbbm{1}}
\newcommand{\zerovac}{\Theta}
\newcommand{\projcont}{\mathbb{P}}

\newcommand{\timeint}{\delta T}

\newcommand{\hmodpsi}{h}

\newcommand{\Top}{\hat{T}}
\newcommand{\Qop}{\hat{Q}}
\newcommand{\Omtr}{\wh{\mathbbm{1}}}
\newcommand{\Omhkll}{\mathbbm{1}}
\newcommand{\Omone}{\Omega_1}

\newcommand{\albarr}{{\cal A}^{\band}}
\newcommand{\albarrfull}{{\cal A}^{\band}_{\text{full}}}
\newcommand{\albarprime}{{\cal B}^{\band}_{\text{full}}}
\newcommand{\alrprime}{{\cal B}^{\diam}_{\text{full}}}
\newcommand{\alr}{{\cal A}^{\diam}}

\newcommand{\alrfull}{{\cal A}^{\diam}_{\text{full}}}
\newcommand{\phihkll}{\phi}
\newcommand{\whphihkll}{{\widetilde{\phi}}}
\newcommand{\tfd}{\Psi_{\text{tfd}}} %Thermofield State
\newcommand{\tfdT}{\Psi_{\text{T}}} %Time shifted thermofield state
\newcommand{\msf}{\mathsf}

\newtheorem{thm}{Theorem}

\def\al{A}

\def\pb[#1,#2]{\{#1, #2\}}
\def\deb[#1,#2]{[#1,#2]_{\text{D.B.}}}

\def\tr{{\rm Tr}}

\def\Or[#1]{{\text{O}}\left({#1}\right)}
\def\dotl[#1,#2]{\left\langle #1,\, #2 \right\rangle}
\def\dotlb[#1,#2]{\left\langle #1,\, #2 \right\rangle}
\def\dotlm[#1,#2]{\left[ #1,\, #2 \right]}
\def\dotp[#1,#2]{(\vect{#1} \cdot\vect{#2})}
\def\aff[#1,#2]{\hat{#1}(#2)}

\def\n4sym{{\cal N}=4 SYM}
\def\>{\rangle}
\def\<{\langle}
\def\weight[#1,#2,#3]{\{(#1),#2,#3\}}
\def\ads[#1]{$\text{AdS}_{#1}$}

\newcommand{\be}{\begin{equation}}
\newcommand{\ee}{\end{equation}}
\newcommand{\ba}{\begin{align}}
\newcommand{\ea}{\end{align}}
\newcommand{\bs}{\begin{split}}
\def\sess\end{split}
\newcommand{\vect}[1]{{\boldsymbol{#1}}}

\def \bea {\begin{eqnarray}}
\def \eea {\end{eqnarray}}
\def \bea* {\begin{eqnarray*}}
\def \eea* {\end{eqnarray*}}

\def \bes {\begin{equation*}}
\def \ees {\end{equation*}}

\def \bandset {{\cal Y}}

\def\alcut[#1]{{\cal A}_{#1, \epsilon}}
\def\alseg[#1,#2]{{\cal B}_{#1, #2}}

\def\projvac{{\cal P}_{\Omega}}
\def\supcharge[#1]{\{#1\}}

\def\projsupeig[#1]{{\cal P}_{{\ell, m}}[{#1}]}
\def\transop[#1, #2]{T_{\{#1\}, \{#2\}}}
\def\supket[#1]{|\{#1\} \rangle}
\def\supbra[#1]{\langle \{#1\} | }

\def\projvac{P_0}

\def\hilbzerosect[#1]{{\cal H}^0_{\{#1\}}}
\def\hilbsect[#1]{{\cal H}_{\{#1\}}}
\def\hilbmass[#1]{{\cal H}^m}

\def\diam{\overline{R}}
\def\band{R}
\def\tband{T_{\text{b}}}
\def\bandset{{\cal B}}
\newcommand{\tone}{\text{I}}
\newcommand{\ttwo}{\text{II}}
\newcommand{\tthr}{\text{III}}

\newcommand{\alg}{\mathcal{A}}
\newcommand{\wh}{\widehat}

\newcommand{\ahkll}{\alg}
\newcommand{\agrav}{\widehat{\alg}}
\newcommand{\hs}{\mathcal{H}}

\newcommand{\beq}{\begin{equation}}
\newcommand{\eeq}{\end{equation}}

\begin{document}

\title{Holographic observers for time-band algebras}
\author[1]{Kristan Jensen,}
\author[2]{Suvrat Raju}
\author[3,4]{and Antony J. Speranza}
\affiliation[1]{Department of Physics and Astronomy, University of Victoria, Victoria, BC V8W 3P6, Canada}
\affiliation[2]{International Centre for Theoretical Sciences, Tata Institute of Fundamental Research, Shivakote, Bengaluru 560089, India}
\affiliation[3]{Institute for Theoretical Physics, University of Amsterdam, Science Park 904, 1098 XH, Amsterdam, The Netherlands }
\affiliation[4]{Department of Physics, University of Illinois, Urbana-Champaign, Urbana IL 61801, USA}
\emailAdd{kristanj@uvic.ca}
\emailAdd{suvrat@icts.res.in}
\emailAdd{asperanz@gmail.com}
\date{}
\abstract{We study the algebra of observables in a time band on the boundary of anti-de Sitter space in a theory of quantum gravity. Strictly speaking this algebra does not have a commutant because products of operators within the time band give rise to operators outside the time band.  However, we show that in a state where the bulk contains a macroscopic observer, it is possible to define a coarse-grained version of this algebra with a non-trivial commutant, and a resolution limited by the observer's characteristics. This algebra acts on a little Hilbert space that describes excitations about the observer's state and time-translated versions of this state.  Our construction requires a choice of dressing that determines how elements of the algebra transform under the Hamiltonian. At leading order in gravitational perturbation theory, and with a specific choice of dressing, our construction reduces to the modular crossed-product described previously in the literature.  We also prove a theorem showing that this is the only crossed product of a type III$_1$ algebra resulting in an algebra
with a trace.  This trace can be used to define entropy differences between states in the little Hilbert space that are insensitive to the properties of the observer. We discuss some technical challenges in extending this construction to higher orders in perturbation theory.  Lastly, we review the construction of interior operators in the eternal black hole and show that they can be written as elements of a crossed product algebra.
}

\setcounter{tocdepth}{1}

\maketitle

%--------------------------------
\section{Introduction}
%--------------------------------

Quantum information measures, such as the entanglement entropy, are commonly studied in the context of many-body quantum mechanics, where one has a well-defined notion of a density matrix for any subsystem and a trace for operators. In the language of von Neumann algebras \cite{murray1936rings,murray1937rings,neumann1940rings,murray1943rings},\footnote{For recent introductions to the subject of von Neumann algebras suitable for high energy physicists, see~\cite{Witten:2018zxz} or \cite{Sorce:2023fdx}. A more mathematical perspective may be found in \cite{Haag:1992hx}.} the algebra of observables in this setting is of type I. In contrast, UV divergences prevent us from associating a finite density matrix or defining a trace on the algebra of observables for a subregion in quantum field theory.  These algebras are said to be of type III and also have been shown to emerge in large $N$ holographic theories
\cite{Leutheusser:2021frk, Leutheusser:2022bgi}.  Recently, there has been interest in the idea that \cite{Witten:2021unn, Chandrasekaran:2022cip} that perturbative gravitational corrections might naturally regulate these UV divergences and improve matters so that one can at least define a trace and entropy differences. These gravitationally-improved algebras form a structure that is called a crossed product and are classified as being of type II.

Motivated by this, the authors of \cite{Jensen:2023yxy} studied the algebra for general subregions in a theory of semiclassical gravity. One case of particular interest is where the region $\diam$ is a ball-shaped region near the middle of anti-de Sitter (AdS) space with a complement $\band$ (see Fig.\ \ref{diamondfig}).  In the approximation where the bulk theory is a local quantum field theory, the algebra of $\band$ is dual to the algebra of a time band on the boundary of AdS and its commutant is dual to the algebra of $\diam$, both of which are type III algebras.  It was argued that, in the presence of an observer, gravitational corrections would convert both algebras to type II algebras. See also~\cite{AliAhmad:2023etg,Klinger:2023tgi} for related discussions on the conversion of algebras from type III to type II.

However, this result leads to a puzzle. In the presence of dynamical gravity, the ADM Hamiltonian is an element of the algebra of operators on any time slice of the boundary. Bounded functions of the Hamiltonian can be used to evolve operators out of the time band via
\be
\label{admhamact}
e^{i H s} O(t) e^{-i H s} = O(t + s)\,.
\ee
A well-defined algebra must be closed under products and include the evolved operator $O(t+s)$. But this suggests that the algebra of the time band cannot be restricted to a finite range but must be extended to the entire boundary, in which case the complementary diamond, $\diam$, disappears together with its algebra.

Said another way, it is possible to ``see'' into the bulk by evolving operators outside the time band.  The standard HKLL reconstruction \cite{Hamilton:2007wj,Hamilton:2006fh,Hamilton:2006az,Hamilton:2005ju} relates quantum field operators at all points in AdS to the integral of light primary operators on the boundary over the light-crossing time of AdS.  Therefore specifying data about light primary operators on the entire boundary is equivalent to specifying data about quantum fields everywhere in the bulk.  This would suggest that the algebra of a time band on the boundary should have no commutant.

This puzzle is closely related to the observation that holography is implicit in the properties of gravity \cite{Marolf:2008mf, Marolf:2013iba,Banerjee:2016mhh,Laddha:2020kvp,Jacobson:2019gnm,Chowdhury:2020hse,Raju:2020smc,Chowdhury:2021nxw,Raju:2021lwh,Chakravarty:2023cll,deMelloKoch:2022sul,deMelloKoch:2024juz}. A careful consideration of the gravitational constraints suggests that observables in a bounded region of spacetime can be approximated arbitrarily well by observables in its complement --- this is called the principle of holography of information.  In the present case, this would suggest that there is a subtlety in associating commuting algebras for a bulk causal diamond and its complement in a gravitational theory.

In this paper, we study and resolve this tension. Our analysis establishes a hitherto unremarked-upon link between recent work on von Neumann algebras and older constructions of approximately local operators in gravitational theories. This sheds light on both sides of the connection. On the algebraic side, this study clarifies the precise role played by the ``observer'' in recent studies of von Neumann algebras, in particular~\cite{Chandrasekaran:2022cip,Jensen:2023yxy}. On the other hand, this study also elucidates algebraic aspects of techniques previously used to explore bulk locality.

Consider a state in AdS that is heavy enough so that the density of nearby states scales exponentially with a (possibly sublinear) power of $N$ --- the ratio of the Planck scale to the cosmological scale.  It is possible for such a state to be much lighter than $N$ so that it does not back react on the classical geometry. This state plays the role of an observer. 

It has long been known that, in the presence of a macroscopic background, it is possible to construct  observables in gravitational theories that behave like approximately local operators in a QFT. (See for instance  \cite{dewitt1960quantization,DeWitt:1967yk,Brown:1994py,Kuchar:1990vy,Sen:2002qa,Giddings:2005id,Marolf:2015jha}.)  Recently --- using techniques developed in \cite{Papadodimas:2013wnh, Papadodimas:2013jku} and refined in \cite{Papadodimas:2015xma,Papadodimas:2015jra}  --- this construction was extended to the case where the state has no distinguishing macroscopic features  \cite{Bahiru:2023zlc,Bahiru:2022oas}.

The main idea is that, under certain conditions, time-translating the observer's state yields a family of almost orthogonal states. This allows for the definition of an independent time operator that can be separated from the time on the boundary. One can then deform the original operators of the theory so that they are naturally parameterized by the new time coordinate.

By adapting these techniques, we define a closed algebra  of observables corresponding to the time band that has a nontrivial commutant.  The operators in this algebra are obtained by deforming the original light-primary operators on the boundary so that the ADM Hamiltonian no longer acts via \eqref{admhamact} but instead acts as a modular transformation on the algebra of the time band.

The formulas that appear in \cite{Papadodimas:2015xma,Papadodimas:2015jra, Bahiru:2022oas,Bahiru:2023zlc} and are used here, at first sight, appear very different from those that appear in the crossed product construction, which leads to type II algebras in theories of gravity, and has been discussed extensively in the recent literature ~\cite{Chandrasekaran:2022cip,Witten:2021unn,Chandrasekaran:2022eqq, Jensen:2023yxy, Kudler-Flam2023, Faulkner:2024gst}.
Remarkably, we find that when we restrict ourselves to leading nontrivial order in perturbation theory and make some specific choices in our construction, the formulas that appear in the literature on bulk reconstruction go over {\em precisely} to  those of  the algebraic literature.
 In particular, the algebra for the time band and its commutant are unitarily equivalent to the algebras for the region $\band$ and $\diam$ described in \cite{Jensen:2023yxy}.

This construction is {\em not} in contradiction with the principle of holography of information and helps to clarify an important physical point. The holographic properties of gravity are analogous to unitarity. Unitarity is an important structural property of any quantum theory and is relevant if one asks precise nonperturbative questions about the reversibility of time evolution. Second, for simple observables, such as perturbative $S$-matrix elements, unitarity can be verified explicitly. However, in a third regime, when one studies physics about a complicated background and coarse-grains the set of observables --- a common setting in everyday life --- physics appears to be dissipative and unitarity is obscured.

The same three regimes apply to the holographic principle.  If one asks precise nonperturbative questions --- for example, about whether information is lost in black-hole evaporation --- the holographic nature of gravity  ensures that the outgoing Hawking radiation contains information about the infalling matter \cite{Raju:2020smc,Raju:2021lwh}. Similarly, if one studies low-energy excitations about the vacuum of AdS or flat space, it is possible to show through explicit perturbative computations that observables in the complement of a bounded region contain information about the region \cite{Chowdhury:2020hse,Chowdhury:2022wcv,Gaddam:2024mqm}. Nevertheless, in a third regime where one introduces a heavy observer and coarse-grains the set of allowed observations, gravitational physics appears to be effectively local and holography is not apparent.

It is commonly believed that the type-II construction in the bulk can be extended to all orders in perturbation theory. This is parallel to the lore that the construction of bulk local algebras that are dressed to the observer should be possible at all orders in perturbation theory. But the technical details of such a construction have never been elaborated explicitly. We list some of the technical challenges that must be addressed to obtain such a higher-order extension. These challenges do not seem insurmountable and  offer avenues for future work.

Our approach also helps to clarify the physical interpretation of the entropy that has been studied in the recent type-II constructions. We show that this entropy corresponds to a coarse-grained entropy that measures the properties of effectively local experiments that can be performed by a bulk observer. As such, these entropies are insensitive to the details of UV-physics or the density of microstates in the UV-complete theory. 

We also find an interesting ambiguity in the construction of our algebras. While the ADM Hamiltonian no longer acts on operators in the algebra via \eqref{admhamact}, its precise action is determined by how we choose to dress the operators. We focus on a choice of dressing where boundary Hamiltonian flow acts on these operators as a version of modular flow and the resultant algebra is a modular crossed product. Other choices of dressing are possible and would yield different crossed products.  From the bulk perspective, the choice to dress the operators via the modular crossed product is motivated by choices of a bulk gauge fixing in the presence of the observer where  time translations correspond to the local boost about the entangling surface. According to the proposed geometric modular flow conjecture \cite{Jensen:2023yxy}, the local boost flow should correspond to modular flow on the bulk quantum fields,  thereby establishing a link between modular dressing and the boost-time gauge-fixing.  

An alternative justification for this choice of dressing is that it is the only one leading to an algebra with a trace.   To demonstrate this, we prove a theorem in section  \ref{sectrmodular} showing that the only crossed product of a type $\tthr_1$ factor resulting in a semifinite algebra  with a well-defined trace is a modular crossed product.  This theorem is a consequence of Takesaki's structure theorem for type $\tthr$ algebras \cite{Takesaki1973}\cite[Theorem XII.1.1]{takesaki2003theory}, but does not appear to have been previously stated in this explicit form.  

The link between relational observables and the crossed product was also noted in the interesting analysis of \cite{DeVuyst:2024pop} that uses the language of quantum reference frames (see also \cite{DeVuyst:2024grw,Chataignier:2024eil,
AliAhmad:2024wja, Fewster:2024pur,
DeVuyst:2024uvd}) and, in a cosmological setting in \cite{Chen:2024rpx,Kudler-Flam:2024psh,Kaplan:2024xyk}.  In our work, the observer is not defined in terms of auxiliary degrees of freedom, or an external particle as in \cite{Chandrasekaran:2022cip}.  Rather the observer and a notion of the observer's time emerges automatically from the properties of the background state,
analogous to the situation
encountered in \cite{Chen:2024rpx,Geng:2024dbl}. Moreover,  the techniques that we use to dress operators  draw upon a parallel stream in the literature \cite{Papadodimas:2015xma,Papadodimas:2015jra,Bahiru:2022oas,Bahiru:2023zlc}.

An overview of this paper is as follows. In section \ref{secpuzzle}, we outline our setup and elaborate on the puzzle alluded to above. In section \ref{secobserver}, we introduce the observer and describe a construction of an extended Hilbert space that describes all coarse-grained low energy experiments that can be performed by the observer. Section \ref{secalgconst} presents some of our main results. We construct the algebra of the time band and its commutant and show that it is unitarily equivalent to the crossed-product presented in \cite{Jensen:2023yxy}.   In section \ref{secentropy} we construct a trace on this Hilbert space and explore the entropy of the states in this effective Hilbert space. In section \ref{sectrmodular} we show how demanding a trace on the algebra necessitates a certain choice of dressing.  In section \ref{sechigherorder} we discuss higher-order and nonperturbative issues. Section \ref{secdiscussion} summarizes our results and outlines future directions. Appendix \ref{seceternalinterior}, which  is out of the main line of this paper, shows how older constructions of interior operators in the eternal black hole can be naturally expressed in the form of a crossed product.

\section{Setup and a potential puzzle }
\label{secpuzzle}

In this paper we study theories of gravity coupled to a small number of light fields in AdS. The details of the UV-completion of these theories will not play a role.

We set the AdS radius to $1$ and consider matter and gravity fluctuations about global AdS$_{d+1}$ written in the coordinate system
\be
ds^2 = -(1 + r^2) d t^2 + \frac{d r^2}{ 1 + r^2} + r^2 d \Omega^2.
\ee
For our construction, it is only important that the bulk geometry should be asymptotically AdS. We choose to work with a geometry that is everywhere AdS for simplicity but our construction can be  generalized to other asymptotically AdS geometries.

We imagine that the bulk contains a matter field, $\phi$, of mass $m$ that is weakly coupled to gravity. It is natural to define an asymptotic limit of this field via
\be
\label{extrapolatedict}
\lim_{r \rightarrow \infty} r^{\Delta} \phi(r, t, \Omega) = O(t, \Omega),
\ee
with $\Delta (\Delta - d) = m^2$. The operator $O(t, \Omega)$, which is simply the boundary limit of a bulk field, is gauge invariant since it does not change under small diffeomorphisms. So it is a natural operator to study in a gravitational theory.

By the standard AdS/CFT dictionary \cite{Maldacena:1997re,Gubser:1998bc,Witten:1998qj}, $O(t,\Omega)$ is a light primary operator in the CFT dual to the theory of quantum gravity in the bulk. However, the discussion in this paper is framed using the perspective of the bulk theory and its asymptotic algebras and does not refer to any specific property of the boundary CFT.

In the coordinates above, the light crossing time of AdS is $\pi$. The objective of this paper is to study the algebra of observables for a time band on the boundary that stretches between $-{\tband \over 2}$ and ${\tband \over 2}$ where $\tband < \pi$. For future reference we define the set of points on the boundary
\be
\bandset = \{(t, \Omega): -{\tband \over 2} \leq t \leq {\tband \over 2} \},
\ee
with no restriction on $\Omega$. We would also like to understand the algebra of operators localized within $\bandset$ and its commutant.  Since the band, $\bandset$, runs across the entire boundary sphere we will sometimes suppress the $\Omega$ coordinate to lighten the notation.

\begin{figure}
\begin{center}
\includegraphics[height=0.4\textheight]{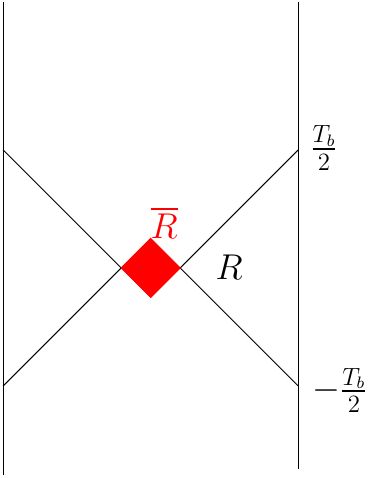}
\caption{\em A boundary time band of extent $\tband$ on the boundary of AdS. The region $\band$ comprises points that can be connected to the band by a timelike curve. Its complement $\diam$ comprises points that are spacelike to the band and is a causal diamond about the middle of AdS.  \label{diamondfig}}
\end{center}
\end{figure}

This description only makes reference to asymptotic operators and so it does not depend on the choice of any bulk coordinate system. Nevertheless, it is useful to think of the bulk geometric setup that is shown in Figure \ref{diamondfig}.
  The set of points that are spacelike separated from this time band form a causal diamond near the middle of AdS. We denote this diamond by $\diam$ and its causal complement by $\band$. 

\subsection{Quantum field theory analysis}
\label{subsecqft}

Before we proceed to gravity, we first remind the reader how the algebra of the time band and its commutant would be defined 
when the bulk is a local quantum field theory in a fixed spacetime. This algebra was first studied in \cite{Banerjee:2016mhh}.

Boundary operators are still defined via the relation \eqref{extrapolatedict}, and we study polynomials of smeared operators in the time band,
\be
\alqft = \text{span}\{O(f_1), O(f_1) O(f_2), \ldots \},
\ee
where
\be
O(f_i) =  \int dt \, d\Omega \, O(t, \Omega) f_i(t, \Omega),
\ee
and $f_i(t, \Omega)$ are functions with support in $\bandset$.  Bounded functions of the polynomial operators above lead to a von Neumann algebra of operators in $\bandset$.

The algebra so obtained is a proper subalgebra of the full algebra of operators. In a quantum field theory, we do {\em not} get all operators in the theory by restricting attention to a time band on the boundary.   Operators at later time on the boundary cannot be obtained by autonomously evolving operators within the time band.

In the bulk the algebra of the time band corresponds to the algebra of bulk quantum fields in the region $\band$.  This is in the spirit of the timelike-tube theorem \cite{araki1963generalization,borchers1961vollstandigkeit,Strohmaier:2023opz} since the region $\band$ comprises the causally dependent region of $\bandset$.

When the bulk is a free-field theory, this correspondence can also be made explicit as follows. Define modes of the boundary operator via
\be
O_{k, l} = {1 \over \tband} \int d\Omega\int_{-{\tband \over 2}}^{{\tband \over 2}}  dt\,O(t, \Omega) Y_{\ell}^*(\Omega) e^{i 2 \pi k {t \over \tband}},
\ee
where $Y_{\ell}(\Omega)$ are spherical harmonics and we have suppressed the magnetic quantum number. Then the bulk field can be written as
\be
\label{freetimehkll}
\phihkll(t, r, \Omega) = \sum_{k, \ell} O_{k, \ell} e^{-i 2 \pi k {t \over \tband}} \zeta_{k, \ell}(r) Y_{\ell}(\Omega) + \text{h.c.},
\ee
where the modes $\zeta_{k, \ell}(r)$ were found explicitly in \cite{Banerjee:2016mhh}. This can be thought of as a mode-version of the HKLL \cite{Hamilton:2007wj,Hamilton:2006fh,Hamilton:2006az,Hamilton:2005ju} bulk reconstruction formula as applied to a boundary time band.\footnote{It is subtle to transform this formula back into position space because, in this case,  the position space HKLL kernel is a distribution that only acts on an appropriate class of operator-valued-distributions. This is discussed further in \cite{Banerjee:2016mhh}. This issue is not relevant here.}

Consequently, the polynomial algebra above can equivalently be written in terms of bulk-smeared fields as
\be
\label{qftalg}
\alqft = \text{span}\{\phihkll(g_1), \phihkll(g_1) \phihkll(g_2), \ldots \},
\ee
where
\be
\phihkll(g_i) = \int dt\, dr \, d\Omega\,\phihkll(t, r, \Omega) g (t,r, \Omega),
\ee
and $g(t, r,\Omega)$ has support only in $\band$.
As above, a von Neumann algebra is constructed through bounded functions of these polynomials.

In the nongravitational theory, the commutant of this algebra corresponds to the bulk quantum field algebra in the complementary diamond $\diam$. This algebra is conveniently described through the full HKLL reconstruction formula. 
\be
\label{freefullhkll}
\phihkll(t, r, \Omega) = \int d\Omega'\int_{-{\pi \over 2}}^{\pi \over 2} d t'  \, K(r, \Omega, \Omega',t, t') O(t', \Omega'),
\ee
where $K$ is the HKLL kernel. The integral here runs over the light-crossing time of AdS and not just the time band. We can now study
\be
\label{commutantqft}
\albarqft = \text{span}\{\phihkll(h_1), \phihkll(h_1) \phihkll(h_2), \ldots \},
\ee
where the smearing is defined just as above but the functions $h_i(t,r, \Omega)$ have support only in the region $\diam$. It can be shown that these operators commute with all elements of $\alqft$.

If one considers nongravitational interactions in the bulk, the formulas \eqref{freetimehkll} and \eqref{freefullhkll} need to be corrected by the addition of double-trace operators on the right hand side \cite{Kabat:2011rz,Kabat:2012av}. The timelike tube theorem suggests that, in the absence of gravity, the algebra of the time band continues to correspond to the algebra of $\band$. Microcausality in the bulk, which is manifest in the absence of gravity, tells us that its commutant continues to correspond to $\diam$.

Since these algebras correspond to bulk subregions it is clear that they are both type III$_1$ von Neumann algebras \cite{Leutheusser:2021frk,Leutheusser:2022bgi}. This may be demonstrated more formally by considering modes near the interface between $\diam$ and $\band$ and showing that one can extract an infinite number of modes that are ``thermally'' entangled \cite{Papadodimas:2019msp}. 

In this paper, we would like to understand these algebras in a theory of gravity. Naively, one might have thought that the definition of these algebras could simply be extended to gravitational theories, order by order, in perturbation theory as can be done in the nongravitational interacting theory. However, as we now explain there is an obstacle to such a construction that appears at leading nontrivial order in the gravitational coupling. 

\subsection{A puzzle}

In the presence of dynamical gravity, it is necessary to include the fluctuations of the transverse traceless gravitons in the bulk and their boundary limits as additional operators in the time band. This is easily done by generalizing the formula \eqref{extrapolatedict} and then generalizing the construction described in section \ref{subsecqft}. Since these dynamical gravitons do not lead to any conceptually new issues, we will continue to work with the prototypical scalar matter described above.

The key difference between gravity and a local quantum field theory arises from the constraints of a gravitational theory. 
These constraints imply that the full Hamiltonian of the theory is an element of the time-band algebra. In the bulk, the energy of the state is measured by the ADM Hamiltonian \cite{Arnowitt:1962hi} that, in Fefferman Graham gauge, is given by \cite{Balasubramanian:1999re,Skenderis:2000in}
\be
\label{admham}
H = {d \over 16 \pi G_N} \lim_{r \rightarrow \infty} \int  d \Omega \, r^{d-2} h_{t t},
\ee
where $h_{t t}$ is the metric fluctuation in the bulk. This is also clear from the holographic dual when it exists. In that case, 
one of the light operators is the dual  stress tensor and the Hamiltonian is simply 
\be
H = \int d \Omega \, T_{t t}.
\ee

These formulas simply reflect the bulk Gauss' law.  In the quantum theory, the implications of the Gauss' law are more profound than in the classical theory. It has been argued \cite{Marolf:2013iba,Marolf:2008mf,Banerjee:2016mhh,Laddha:2020kvp,Jacobson:2019gnm,Chowdhury:2020hse,Raju:2020smc,Chowdhury:2021nxw,Raju:2021lwh,Chakravarty:2023cll,deMelloKoch:2022sul,deMelloKoch:2024juz}  that gravitational theories are holographic precisely because the Hamiltonian is a boundary term.  These arguments have implications both in the perturbative and the nonperturbative regime.

For our purposes, we only note that since the Hamiltonian is an element of the time band, repeating the construction above does not lead to a nontrivial subalgebra. An operator outside the time band can be written in terms of operators in the time band via
\be
\label{timeevolve}
O(t, \Omega) = e^{i H t} O(0, \Omega) e^{-i H t}.
\ee
An algebra must be closed under products. Since all operators on the right hand side are elements of the algebra, applying the formula above for $|t| > {\tband \over 2}$ tells us that the closure of the time-band algebra comprises all operators on the entire AdS boundary.

We would like to make a few comments. First, one might be concerned that the boundary operator defined by \eqref{admham} does not generate time translations, and so one might be suspicious of formula \eqref{timeevolve}. In section \ref{sechigherorder} we provide another, more physical argument, that simply relies on a physical measurement of the energy at the boundary and shows that operators in the time band comprise all operators at the boundary.  

Second, practitioners of quantum field theory are often at home with polynomial algebras of light operators. From this perspective, while it is clear that the Hamiltonian must be included in the set of observables at leading nontrivial order in $G_N$, one might be concerned that $e^{-i H t}$ involves very high powers of the Hamiltonian when it is expanded in a power series. But von Neumann algebras are defined in terms of bounded functions of smeared local operators and from that perspective it is natural to study the time-evolution operator.

\subsection{Physics of the resolution}

This tension outlined above is physically important and its resolution does not rely on a technicality but involves interesting physics. It turns out there are some regimes of interest in which the definition of the time-band algebra does not make sense and some regimes in which it does.

There are two regimes in which one should not think of the algebra of the time band as a proper subalgebra of the full algebra.
\begin{enumerate}
\item
In some problems, such as the black-hole information paradox, nonperturbatively small effects are important. A careful examination of the constraints in the quantum theory provides a mechanism for information to leak out from within the black-hole interior invalidating Hawking's argument for information loss \cite{Raju:2020smc}. 

If one considers a black hole in AdS then an observer making exponentially precise measurements in a time band on the boundary can always verify that the state is pure and that information has not been lost.

The question of whether information is lost is an inherently fine grained question since pure states are exponentially close to mixed states: there is no sense in restricting the bulk observer to coarse-grained observations and then asking questions about unitarity or information loss.
\item
In a different regime, when one focuses on low-energy states, it was shown in \cite{Chowdhury:2020hse} that observers  on the boundary could perform a set of physical operations within perturbation theory --- corresponding to turning on a local unitary followed by a measurement of the energy --- to detect the state of the bulk. 

Therefore, the set of observables in the time band, when studied about perturbative states, has no commutant. This includes the case of the AdS vacuum.
\end{enumerate}

Nevertheless, there is a {\em third regime} that can be studied in addition to the two regimes described above.  This third regime is characterized by a specific kind of background state and by limitations on the possible observations.
\begin{enumerate}
\item
Instead of studying fluctuations about empty AdS, we study fluctuations about a heavy background state. This background can be thought of as corresponding to an observer. We will consider a background state that is parametrically heavier than the AdS scale but much lighter than the Planck scale. The presence of this background state takes us out of the regime of validity of the perturbative protocol described in  \cite{Chowdhury:2020hse}. 
\item
Second, we coarse-grain observations on the time band. This coarse-graining takes us out of the regime of validity of \cite{Laddha:2020kvp}. In this context, the coarse-graining is acceptable since we are not interested in delicate questions such as the purity of the full state including the background but merely in describing a set of experiments performed by an observer with limited accuracy.
\end{enumerate}
The objective of this paper is to describe this regime in some detail. We will show that, in this regime, it makes sense to study the algebra of a time-band. Moreover, the algebra has a well-defined commutant.

\section{Observer and extended Hilbert space }\label{secobserver}

\subsection{The observer}

In this paper, the observer is simply a heavy state, $|\obsst \rangle$, in the full theory. This state must satisfy several physical conditions that we outline in this section.

First we choose the state so that its energy
\be
\obsen = \langle \obsst | H |\obsst \rangle,
\ee
satisfies
\be
1 \ll \obsen   \ll N,
\ee
where the AdS scale has been set to $1$ and $N$ is used to denote the Planck scale, where backreaction becomes important. The inequality $\obsen \ll N$ ensures that the state does not backreact significantly on the geometry, which can still be taken to be global AdS to a good approximation. As mentioned above, there is no obstacle in generalizing this construction to the case where the observer does backreact and the geometry is only asymptotically AdS. We choose the regime above only for simplicity.

We need the spread in energy of the observer to be large compared to the AdS scale,
\be
(\obsenfluct)^2 = \langle \obsst | H^2 |\obsst \rangle - \langle \obsst | H | \obsst \rangle^2 \gg 1\,.
\ee
We also need the spread of energies to be much smaller than the energy itself.
\be
\obsenfluct \ll E.
\ee
Physically,  this spread allows
the observer to measure time intervals $\timeint$ satisfying
\be
\timeint \gg {1 \over \obsenfluct}.
\ee

The logarithm of the number of states in the energy range, $E \pm \obsenfluct$  is denoted by $S$ and as a consequence of the assumption above, we also have
\be
1 \ll S \ll N.
\ee
We take the state $|\obsst \rangle$ to be a linear combination of these $e^{S}$ states with random coefficients.

If we lived in a world where the cosmological constant had the same value as it does today but the opposite sign,  we could take the observer to be a state with a single virus on top of the vacuum.  A virus with a mass of  $10^{-15} {\rm g}$ \cite{bahr1976determination} is much lighter than the Planck scale of $2 \times 10^{-5} {\rm g}$ but much heavier than the cosmological scale, which is only $4 \times 10^{-66} {\rm g}$.  
We list some additional required properties of the observer  in section \ref{subsecprop}.

\subsection{Coarse-grained sets of observables}

We define a cutoff set of polynomials in light primary operators in the time band, while excluding the Hamiltonian,
\be
\label{altcdef}
\altc = \text{span}\{O(f_1), O(f_1) O(f_2), \ldots, O(f_1) O(f_2) \ldots O(f_n) \},
\ee
where $f_i$ are some smooth functions that have support in the time band i.e. 
\be
f_i(t, \Omega) = 0, \quad \text{if}~ |t| > {\tband \over 2}.
\ee
We cut off the polynomials at some value 
\be
n \ll S.
\ee

Two points require explanation. First, the difference between $\altc$ as defined in \eqref{altcdef} and $\al_{\text{QFT}}$ is that, at this point, $\altc$ is just a set. It is not an algebra because of the cutoff that is imposed on the size of its elements. Due to this cutoff, the set is not closed under products. 
Second, the exclusion of the Hamiltonian from the set \eqref{altcdef} is a subtle point. In fact,  the OPE of two light operators can generate the stress tensor on the boundary and, therefore, the Hamiltonian. At leading order in the ${1 \over N}$ expansion, where we neglect this effect, it makes sense to separate  light boundary operators and the Hamiltonian. We discuss this point further in section \ref{sechigherorder}. 

Later in the paper, both these constraints will be relaxed: we will take $N, S,$ and $ n$ to infinity and also include the Hamiltonian in the set of allowed observables. This leads to an algebra but this limit must be taken carefully as explained below.

\subsection{Little Hilbert spaces}

The presence of the observer allows us to construct the following little Hilbert space:
\be
{\cal H}_0 = \altc |\obsst \rangle.
\ee
This is the space of states constructed by acting with all elements of $\altc$ on $|\obsst \rangle$. ${\cal H}_0$ inherits a natural vector-space structure from $\alset$. 
Physically one should think of ${\cal H}_0$ as the natural space to ask questions about simple experiments performed in the state $|\obsst \rangle$.

The construction of the space here is precisely the same as the little Hilbert space introduced originally in \cite{Papadodimas:2013jku,Papadodimas:2013wnh}. We need the elements of ${\cal H}_0$ to be normalizable. This does not require us to use bounded functions of the operators in $\altc$. Rather, it is sufficient to slightly smear the field operators in $\altc$ so as to obtain normalizable states in ${\cal H}_0$.

Recall that, as discussed above, the set $\altc$ does not contain the Hamiltonian. So we may consider a family of little Hilbert spaces labelled by a parameter $t$ obtained by time-evolving the original space by $t$.
\be
\label{htdef}
{\cal H}_{t} = \altc  e^{-i H t}  |\obsst \rangle.
\ee
These states can be thought of as the observer translated in time and then excited with an element of $\altc$.

\subsection{Additional properties of the observer \label{subsecprop}}

We now place some additional, physically motivated,  restrictions on the observer that will help in simplifying the structure of the little Hilbert spaces described above. 
Roughly speaking, the objective of these conditions 
is to ensure that the observer acts as a non-interacting thermal bath.

First, we demand that from the point of view of the small algebra of operators time-translated states of the observer appear to be the same as the original observer state. This means that
\be
\label{timetransinv}
\langle \obsst | e^{i H t} a e^{-i H t} |\obsst \rangle = \langle \obsst | a |\obsst \rangle, \qquad \forall a \in \altc.
\ee
It is important that $e^{-i H t} |\obsst \rangle \neq |\obsst \rangle$ and the criterion above only indicates the equality of expectation values to leading order in $N$.  This criterion is quite easy to achieve. If we take the observer to be a superposition of $e^{S}$ energy eigenstates with the spread $\obsenfluct$ 
\be
|\obsst \rangle = \sum_{i} c_i |E_i \rangle,
\ee
then 
\be
\label{timeinvexpansion}
\langle \obsst | e^{i H t} a e^{-i H t} | \obsst \rangle = \sum_{i} \langle E_i | a |E_i \rangle |c_i|^2 + \sum_{i \neq j} \langle E_i | a | E_j \rangle c_i^* c_j e^{i (E_i - E_j) t}.
\ee
The first term is manifestly time-independent. The second term comprises cross terms that tend to average out to zero unless the state and the operator-insertion are finely attuned to another. For a generic state, the fluctuations induced by the second term can be estimated to be $\Or[e^{-{S \over 2}}]$ \cite{lloyd1988black}.

The requirement \eqref{timetransinv} means that we are in a slightly different setup from the one usually considered in defining quasilocal bulk operators. In those setups, one often studies a background that has macroscopic time dependence. We will see that \eqref{timetransinv} leads to a simplified structure for the effective Hilbert space and is useful to obtain a crossed product structure for the final algebra.
  
Second, the observer's state must satisfy some energy conditions. Physically, we require that (a) the energy distribution of the observer fall off sufficiently rapidly at high energies and (b) that it be very difficult for a simple probe drawn from $\altc$  to extract a large amount of energy from the observer.
These can be framed mathematically as the requirement that for all elements $a$, and $b$ of $\altc$  the correlator
\be
\label{analcontmatrix}
\langle \obsst | b e^{-s H} a e^{s H} |\obsst \rangle,
\ee
is convergent for $0 < \text{Re}(s) < \alpha$ for some value of $\alpha$, independent of $a$ and $b$. 

Provided that the insertion of $b$ cannot significantly lower the energy of the state $|\obsst \rangle$, the insertion of $e^{-s H}$ provides a damping factor that offsets the factor provided by $e^{s H}$ on the right. 

To understand this criterion better, consider the analogous correlators in the vacuum and in a thermal state. In the vacuum, we have
\be
\langle \Omega | b e^{-s H} a e^{s H} | \Omega \rangle = \sum_{E} \langle \Omega | b | E \rangle \langle E | a | \Omega \rangle e^{-s E},
\ee
where we have inserted a complete set of energy eigenstates in the correlator and used the property $H | \Omega \rangle = 0$. Since the energy is always positive we see that the factor that was inserted always improves the convergence of the sum over $E$ provided $\text{Re}(s)$ is positive. Therefore, in the vacuum, this correlator is convergent for all positive $\text{Re}(s)$.

Next, consider a thermal state at temperature $\beta^{-1}$. In such a state we have
\be
\tr(e^{-\beta H} b e^{-s H} a e^{s H}) = \tr(e^{-(\beta - s) H}  b e^{-s H} a),
\ee
using the cyclicity of the trace. Using the positivity of energy, we expect that this correlator converges provided $0 < \text{Re}(s) < \beta$.

To be clear: we are not demanding that correlators in the observer state be precisely like thermal correlators or like vacuum correlators. We only demand that they share the property of vacuum and thermal correlators in allowing the matrix element \eqref{analcontmatrix} to exist.

\subsection{A Reeh-Schlieder-like result}

The conditions imposed upon the observer allow us to prove a Reeh-Schlieder-like result \cite{Haag:1992hx} for the little Hilbert space. Consider the space 
\be
\label{htauprimedef}
{\cal H}_{t}' = \text{span}\{O(g_1), O(g_1) O(g_2), \ldots, O(g_1) O(g_2 ) \ldots O(g_n)  \} e^{-i H t} |\obsst \rangle,
\ee
where $g_i$ have support for {\em all t} that is not restricted to the time band. The  larger domain of the functions $g_i$ distinguishes $H_{t}'$ from $H_{t}$. But we now show that ${\cal H}_{t}$ is dense in ${\cal H}_{t}'$. 

We first prove that  ${\cal H}_0$ is dense in ${\cal H}_0'$. Say there was a state  $|\eta \rangle$ in ${\cal H}_{0}'$ that was orthogonal to all states in ${\cal H}_0$. Then this state must have the property that
\be
\label{wrongassumption}
\langle \eta | O(t_1) \ldots O(t_n) |\obsst \rangle = 0, \qquad \forall t_i \in \left[-{\tband \over 2}\,, {\tband \over 2}\right]\,.
\ee
This means that 
\be
\label{etaorth}
\langle \eta | O(t_1) \ldots e^{i H s} O(t_n) e^{-i H s} | \obsst \rangle = 0\,,
\ee
provided $t_n + s$ is in the time band. 

The condition \eqref{analcontmatrix} implies that \eqref{etaorth} is analytic when $s$ is continued in the strip
\be
\text{Im}(s) < \alpha\,.
\ee
By assumption we know that the correlator in \eqref{etaorth} vanishes when $t_n + s$ is in the time band and the property of analyticity implies that it must vanish for all $s$. Therefore, our assumption leads to the conclusion that \eqref{wrongassumption} vanishes even when we relax the condition that $t_n \in [-{\tband \over 2}, {\tband \over 2}]$. 

We can now repeat the argument by simultaneously shifting both $t_{n-1}$ and $t_n$ by $s$ as explained in \cite{Witten:2018zxz}. Since we are allowed to make independent shifts on $t_n$ without affecting the vanishing of the correlator, this implies that one can move both $t_{n-1}$ and $t_{n}$ outside the time band. Proceeding this way, our assumption leads to the conclusion that \eqref{etaorth} vanishes when $t_i$ are arbitrary. But this is impossible since $|\eta \rangle$ belongs to the space defined in \eqref{htauprimedef}.

We are led to conclude that our assumption  \eqref{wrongassumption} must be false and that $|\eta \rangle$ satisfying \eqref{wrongassumption} cannot exist.  Therefore, there is no element of ${\cal H}_{0}'$ that is orthogonal all elements of ${\cal H}_{0}$.  Since ${\cal H}_{0}$ is clearly a subset of ${\cal H}_{0}'$, this implies that ${\cal H}_{0}$ is dense in ${\cal H}_{0}'$. 

It is trivial to generalize the result above to show that ${\cal H}_t$ is dense in ${\cal H}_{t}'$. A simple corollary of these results is the following.  Consider
\be
\label{htotherorderdef}
{\cal H}_{t}'' = e^{-i H t} \altc  |\obsst \rangle.
\ee
Then the result above shows that ${\cal H}_{t}$ is dense in ${\cal H}_{t}''$ and vice versa.

\subsection{Overlaps}

Now consider the overlap of ${\cal H}_{t}$ with ${\cal H}_{0}$. We see that with $a \in \altc$
\be
\label{overlapexp}
\langle \obsst | a e^{-i H t} | \obsst \rangle = \sum_{i}  |c_i|^2 \langle E_i | a | E_i \rangle e^{-i E_i t} + \sum_{i \neq j} c_i^* c_j e^{-i E_j t} \langle E_i | a | E_j \rangle.
\ee
The difference with \eqref{timeinvexpansion} is that both terms now vary with $t$.

In the limit where $t \gg {1 \over \obsenfluct}$, the action of $e^{-i H t}$ induces random phases in the correlator. Then a simple random-phase estimate suggests that we have
\be
\label{overlapfalloff}
|\langle \obsst | a  e^{-i H t}  |\obsst \rangle | \ll 1, \qquad t \gg {1 \over \obsenfluct}.
\ee

This result crucially relies on the exponentially large density of states in the vicinity of the original observer state and the spread in energies about this state. The result above would not hold if we had taken $|\obsst \rangle$ to be the vacuum or a state close to the vacuum, or even a collection of energy eigenstates with only a small spread in energies.

The precise falloff of the overlap in \eqref{overlapfalloff} with $t$ depends on the details of the state and might even depend on the operator $a$ that is inserted. To get rid of this dependence, it is useful to put the time-translated little-Hilbert spaces into bins with a size larger than ${1 \over \obsenfluct}$.

We choose a time-interval
\be
\timeint \gg {1 \over \obsenfluct},
\ee
and define a smeared little Hilbert space
\be
{\cal H}^{\text{sm}}_{t} = \text{span}\left\{|v\rangle: |v\rangle \in {\cal{H}}_{s}~\text{with}~s \in \left(t - {\timeint \over 2}, t + {\timeint \over 2}\right)\right\}.
\ee
The projector onto this space is denoted by $P_{t}$. Alternately, we can think of $P_{t}$ as the smallest projector that contains projections onto all the little Hilbert spaces in the range $(t - {\timeint \over 2}, t + {\timeint \over 2})$.
We have,
\be
|P_{t} P_{t'}| \ll 1 
\ee
for $|t - t'| \gg {1 \over \obsenfluct}$ where $|\cdot|$ is the operator norm.  But these projectors are not orthogonal if this condition is not met. 
We can now define the operator
\be
\label{timedef}
\Top = \sum_{t} t P_{t},
\ee
where the sum is taken over a discretization of the real line in units of $\timeint$ i.e. $t \in \{n\delta T : ~ n \in\mathbb{Z} \}$.

\subsection{Smooth states and enlarged space}\label{subsecenlarged}

We now restrict attention to a set of smooth states that live in the union of the little Hilbert spaces described above on which the time operator will act nicely. Define
\be
\label{fastate}
|f,a \rangle = {\cal N} a \int d t f(t) e^{-i H t}  |\obsst \rangle,
\ee
where $f(t)$ is a smooth function that does {\em not} vary on scales smaller than $\timeint$ and ${\cal N}$ is a normalization factor that we will fix below.  The specification of the state depends both on $f$ and an element of the algebra $a$. The ordering of the operators above is important and the state is defined by first acting with $e^{-i H t}$ and then acting with $a$. We will sometimes be interested in the case $a = \mathbbm{1}$, which we will denote by
\be
\label{f1state}
|\fullst \rangle = {\cal N} \int d t f(t) e^{-i H t} |\obsst \rangle.
\ee

The ordering chosen ensures that the elements of the algebra themselves act simply on the basis above. In particular, we have 
\be
a |f, b \rangle = a b \int d t \,  f(t)  e^{-i H t} |\obsst \rangle = |f, a b \rangle.
\ee
On the other hand, the Hamiltonian does not act in a simple manner.
\be
\label{actham}
e^{-i H s} |f, a \rangle = \int  d t \,e^{-i H s} a f(t) e^{-i H t}  | \obsst \rangle = \int d t \, e^{-i H s} a e^{i H s} e^{-i H (t + s)} f(t)  | \obsst \rangle.
\ee
While the operator $e^{-i H s} a e^{i H s}$ is {\em not} an element of the time band algebra for large enough $s$, we have shown above that its {\em action} on the state on the right does belong to the direct sum of little Hilbert spaces constructed above. Nevertheless, the right hand side of \eqref{actham} cannot be rewritten in the form \eqref{fastate} without knowing more details of correlators in the state $|\obsst \rangle$. 

We will now analyze the action of the time operator on such states and explore the structure of the inner product. We will find simple formulas that are accurate up to corrections suppressed by $\Or[{1 \over \obsenfluct \timeint}]$ and $\Or[e^{-S}]$ and also $\Or[f'(t) \timeint]$. We will collectively denote these corrections as $\Or[\epsilon]$ below. We will shortly take a limit where all these corrections vanish.

The action of the projectors $P_{t}$  on the states above is given by
\be
\label{ptonfa}
P_{t} |f, a \rangle  = {\cal N} f(t) \int_{{t - {\timeint \over 2}}}^{t + {\timeint \over 2}} d t'\,  a e^{-i H t'}  |\obsst \rangle  + \Or[\epsilon].
\ee
The correction term arises because the different $P_{t}$ are not exactly orthogonal projectors. In particular, $P_{t}$ acts nontrivially on some states with $t'$ that is beyond the range of integration.  However, these edge effects are confined an interval ${1 \over \obsenfluct \timeint}$. 
As a consequence of the equation above, we also find that
\be
\Top |f, a \rangle = \int dt\, t f(t) a e^{-i H t}   | \obsst \rangle + \Or[\epsilon].
\ee

Finally, let us study the norm on this space. We find that
\be
\langle g,b | f, a  \rangle =  {\cal N}^2 \int d t d t' \langle \obsst | e^{i H t'} b^{\dagger}  a e^{-i H t}  | \obsst \rangle f(t) g^*(t').
\ee

Now we note that 
\be
 \langle \obsst | e^{i H t'} b^{\dagger}  a e^{-i H t} | \obsst \rangle \approx 0, \qquad \text{when}~|t'-t| > \timeint,
\ee
and also that
\be
f(t) g^*(t') \approx f(t) g^*(t) \qquad \text{when}~|t'-t| < \timeint.
\ee
This allows us to approximate 
\be
\label{normprecise}
\langle g,b | f, a  \rangle =  \Lambda_{b a} \int d t f(t) g^*(t)  + \Or[\epsilon],
\ee
where
\be
\label{lambdaabdef1}
\Lambda_{b a} = {\cal N}^2 \int_{-\timeint}^{\timeint} d s \langle \obsst | e^{i H s} b^{\dagger} a | \obsst \rangle,
\ee
is independent of $t$. Here we have used \eqref{timetransinv}. 

By defining 
\be
|\smobsst \rangle = {{\cal N} \over (\timeint)^{1 \over 2}} \int_{-{\timeint \over 2}}^{{\timeint \over 2}} d s\,  e^{-i H s} |\obsst \rangle,
\ee
we can also write
\be
\label{lambdaabdef}
\Lambda_{b a} =  \langle \smobsst | b^{\dagger} a | \smobsst \rangle.
\ee
Equation \eqref{lambdaabdef} can be seen to be equivalent to \eqref{lambdaabdef1} by expanding the right hand side
\be
\begin{split}
\langle \smobsst | b^{\dagger} a | \smobsst \rangle &= {{\cal N}^2 \over \timeint} \int_{-{\timeint \over 2}}^{\timeint \over 2} d s \, \int_{-{\timeint \over 2}}^{\timeint \over 2} d s' \, \langle \obsst | e^{i H s'} b^{\dagger} a e^{-i H s} | \obsst \rangle \\
&= {{\cal N}^2 \over \timeint} \int_{-{\timeint \over 2}}^{\timeint \over 2} d s \int_{-{\timeint \over 2} - s}^{{\timeint \over 2} - s} d s'' \langle \obsst | e^{i H s''} b^{\dagger} a | \obsst \rangle\,,
\end{split}
\ee
where, by \eqref{timetransinv}, the integrand only depends on  $s''=s'-s$.  Moreover, the precise limits on $s''$ are not important because the integrand falls off rapidly in the range ${1 \over \delta E} \ll |s'-s| \ll \timeint$ yielding \eqref{lambdaabdef1}. So the integral over $s''$ is independent of $s$ and the integral over $s$ just cancels off the factor of $\timeint$ leading to \eqref{lambdaabdef1}. 

The formula \eqref{lambdaabdef} makes it clear that  $\Lambda_{b a}$ is positive definite since $\Lambda_{a a}$ is just the expectation value of a positive operator in the a smeared version of $|\obsst \rangle$.

It is now convenient to choose the normalization factor ${\cal N}$ so that the norm simplifies when we take both $a,b$ to be the identity operator
in the formulas above.
\be
\Lambda_{\mathbbm{1} \mathbbm{1}} = 1.
\ee

\subsection{Limiting Hilbert space} 
\label{subseclimit}

In the discussion above, we discretized the set of observables.
In this paper, we will be interested in the limiting case where the effects of discretization can be neglected. (See \cite{Soni:2023fke} for related discussion.)

The key limit is to take 
\[
N \rightarrow \infty.
\]
In this limit, the gravitational coupling becomes arbitrarily weak although, of course, we still intend to include the Hamiltonian in the algebra and also the states generated by its action. Alongside, we also take 
\[
S \rightarrow \infty, \qquad E \rightarrow \infty, \qquad \delta E \rightarrow \infty.
\]
It is then also possible to take 
\[
n \rightarrow \infty, \qquad \timeint \rightarrow 0.
\]
As we take these limits it is necessary to also ensure that
\[
{n \over S} \rightarrow 0, \qquad {\delta E \over E} \rightarrow 0, \qquad \timeint \delta E \rightarrow \infty.
\]

An alternate way to view the limit procedure above is as follows. We continue to work in the discrete and precise setting above but we neglect small corrections proportional to ${1 \over N}, {1 \over S}, {1 \over n}, {n \over S}, {1 \over \timeint \delta E}$ and ${\delta E \over E}$. 

While the normalization factor ${\cal N}$ above blows up as ${(\delta E)^{1 \over 2}}$ in this limit, note that $\Lambda_{b a}$ as defined in \eqref{lambdaabdef} still remains finite.

For most of the rest of this paper, we will work in this limit. In this limit $\altc$ turns into a closed algebra. This is clear, since in the absence of a cutoff, one can multiply two elements of $\altc$ to get another element. This algebra is of type $\text{III}_1$ \cite{Leutheusser:2021frk, Leutheusser:2022bgi}.

\subsection{Direct product structure}

In the limiting Hilbert space the small terms that appear on the right hand side of \eqref{normprecise} vanish. Moreover, in the discussion above, the functions $f(t)$ were constrained to be vary slowly on the scale $\timeint$ but after taking the limit, this restriction is unimportant. Consequently, the set of states described above takes the form of a direct product 
\be
\label{tensordecomp}
|f, a \rangle \in \honedim \otimes \altc,
\ee
where $\honedim$ is simply the space of $L^2$ normalizable function representing the Hilbert space of a single particle in quantum mechanics.   We are using the algebra itself  as a representation of the other tensor factor of the Hilbert space.

After dropping the $\Or[\epsilon]$ terms above, the norm on this space is
\be
\langle g, b | f, a \rangle = ( g, f ) \langle \smobsst|b^{\dagger} a | \smobsst \rangle,
\ee
where $(g, f)$ denotes the standard $L^2$ norm on the space of functions.

Note that the elements of $\altc$ act only on the second factor in the tensor decomposition \eqref{tensordecomp} whereas the $\Top $ operator acts only on the first factor.

\paragraph{Time eigenstates.}

Going beyond normalizable wavefunctions, one can also consider a set of eigenstates of the $\Top $ operator 
\be
\Top |t_{\Top } \rangle = t |t_{\Top } \rangle.
\ee
These are normalized as
\be
\label{torthog}
\langle t_{\Top } | t'_{\Top } \rangle = \delta(t - t'),
\ee
where the delta function is understood as a distribution that acts on functions that vary slowly over a range $\timeint$.  Within this space of functions, the eigenstates above are complete. 

In this basis, we can also write states in the direct product space above in the form
\[
|t_{\Top }, a \rangle
\]
We will continue to use the $|f, a \rangle$ notation and in that notation the eigenstate above corresponds to $f(s) = \delta(s-t)$. 

It will also be convenient to study the projection-valued measure
\be
\projcont_{t} = |t_{\Top } \rangle \langle t_{\Top }|,
\ee
which acts only on the $\honedim$ factor.  This projector-density should not be conflated with the projectors $P_{t}$ that were introduced before going to the limiting Hilbert space in section \ref{subsecenlarged};
in particular, $\projcont_t^2$ is divergent,
while $P_t^2 = P_t$. In terms of this projector density  we have
\be
\Top = \int dt\,t \projcont_{t}.
\ee
The action of this projector density on normalizable states is given by
\be
\projcont_{t} |f, a \rangle = f(t) |t, a \rangle.
\ee

\paragraph{Canonical conjugate of the time operator.}

The canonical conjugate of $\Top $ will be denoted by $\Qop$ which satisfies
\be
[\Top , \Qop]=i.
\ee
We demand that $\Qop$ commutes with the algebra itself. 
\be
[\Qop, a] = 0.
\ee
Therefore, its action on elements of the algebra is simply
\be
\label{defq}
e^{-i \Qop s} |t_{\Top }, a \rangle = |(t + s)_{\Top }, a \rangle.
\ee
The equation \eqref{defq} can be taken as the {\em definition} of the operator $\Qop$.

This operator will play an important role in what follows. A useful identity to keep in mind is
\be
\label{qshift}
e^{i \Qop s} \Top e^{-i \Qop s} = \Top + s.
\ee
We remind the reader that similar identities hold between $H$ and $\Top $:
\be
\label{hshift}
[\Top , H] = i; \qquad e^{i H s} \Top e^{-i H s} = \Top + s.
\ee
The relation \eqref{defq} does not hold when $\Qop$ is replaced with $H$.  The difference between $\Qop$ and $H$ has to do with their commutator with elements of $\altc$. But these operators commute with each other
\be
[\Qop, H] = 0.
\ee

\paragraph{Vacuum state.}

The norm in the direct product Hilbert space involves expectation values of a product of two operators in the state $|\smobsst \rangle$. Nevertheless the original smeared state  of the observer $|\smobsst \rangle$ does not have a natural representation in the direct product Hilbert space that emerges after taking the limit \ref{subseclimit}.  It is convenient to study the state
\be \label{eqn:Theta}
|\zerovac \rangle = |0_{\Top }, \mathbbm{1} \rangle,
\ee
obtained by taking $f(t) = \delta(t)$ 
in (\ref{f1state}).
$|\zerovac \rangle$ is not normalizable. Nevertheless, it shares some of the properties of $|\smobsst \rangle$ in that (up to $\Or[\epsilon]$ terms that vanish in the limit under consideration)
\be
\Top |\zerovac \rangle = 0; \qquad \Top |\smobsst \rangle = 0.
\ee
While expectation values of elements of $\altc$ in $|\zerovac \rangle$ are ill-defined due to the $\delta(0)$ divergence in the $\honedim$ sector, we have
\be
\label{omegaproptosmst}
\int d t d t'\langle \zerovac | f^*(t) e^{i \Qop t} a f(t') e^{-i \Qop t'}| \zerovac \rangle  = \left(\int d t\, |f(t)|^2 \right) \langle \smobsst | a | \smobsst \rangle.
\ee

\section{Algebra of the time band and its commutant} \label{secalgconst}

In the previous section, we described the construction of an enlarged Hilbert space that captures the action of light operators and also the action of the Hamiltonian on a background state that describes an observer. In this section, use this construction to define an algebra for the time band and the commutant. 

We will always work within the limit described in section \ref{subseclimit}. Although we continue to use the notation $\altc$ to describe the collection of light operators that can act on the observer, this set into an algebra in the limit since we now include arbitrarily many products of operators. This algebra coincides with the algebra of quantum field theory without gravity which, as we have already noted, is type $\text{III}_1$. Our objective now is to consistently include the Hamiltonian in this algebra. We do this through the following steps
\begin{enumerate}
\item
The operators in $\altc$ can be deformed so as to dress them to the observer rather than to the boundary. While the action of the deformed operators on the original observer's state is the same as the action of the undeformed operators, their commutator with the Hamiltonian is altered. This construction slightly generalizes previous work \cite{Papadodimas:2015xma,Papadodimas:2015jra,Bahiru:2022oas,Bahiru:2023zlc} that showed how to alter the commutator of light operators with the Hamiltonian. 
\item
The Hamiltonian can now be included in the algebra of deformed operators, which is closed under its action. Our construction differs slightly from previous constructions because the Hamiltonian does  not commute with the deformed operators described in step 1. Rather the boundary Hamiltonian acts on these operators as an automorphism. 
\item
We explicitly construct the commutant of the algebra obtained via steps 1 and 2.
\item
Finally, we show that with a specific choice of automorphism, the algebra of the time band and its commutant are precisely the same as the crossed-product construction that has appeared in the literature \cite{Jensen:2023yxy}. Our analysis thus establishes a relation between older constructions of relational observables and recent work on the crossed-product.
\end{enumerate}

\subsection{Algebra of the time band}

In this subsection, we will construct the algebra of the time band. We first introduce an automorphism of the algebra $\altc$  that will be used in this construction.  

\subsubsection{Automorphism}

An automorphism of an algebra is an operation that maps the algebra back to itself and preserves its structure \cite{takesaki2006tomita}. Consider an operator $X$ that generates an automorphism of $\altc$ so that for any real $t$,
\be
X^{i t} a X^{-i t} \in \altc, \qquad  \forall a \in \altc.
\ee
This operation preserves the structure of the algebra because 
\be
X^{i t} a b X^{-i t}  = (X^{i t} a X^{-i t}) (X^{i t} b X^{-i t}).
\ee
The definition of the automorphism makes sense after the limit of section \ref{subseclimit} has been taken so that $\altc$ can be considered to be closed under products.

If $X \in \altc$, this automorphism is called an inner automorphism.  If $X \notin \altc$, this automorphism is called an outer automorphism. 

\paragraph{Modular automorphism.}

A prototypical example of an outer automorphism for type $\text{III}_1$ algebras 
 is the modular automorphism. The modular operator is defined as the operator that acts on the $\altc$ factor in \eqref{tensordecomp} and satisfies
\be
\label{modopdef}
\begin{split}
&\langle \fullst | a \Delta b| \fullst \rangle = \langle \fullst | b a | \fullst \rangle.\\
&[\Delta, \Top ] = [\Delta, \Qop] = 0.
\end{split}
\ee
Note that \eqref{modopdef} completely specifies the action of $\Delta$ on all states and so completely determines the operator.
$\Delta$ itself is not in the algebra, $\altc$,  but it may be shown that conjugation by $\Delta^{i t}$ still maps elements of the algebra back to itself \cite{takesaki2003theory}.

In this section, we keep $X$ unspecified. Different choices of $X$ will turn out to correspond to different definitions of the algebra of the time band and its commutant. In the next section, we will restrict $X=\Delta^{-1}$  and show that,  with this choice, the resulting algebra has nice properties and admits a trace. 

\subsubsection{Deformed operators}

Now using the projectors and automorphism introduced above, we define the following hatted operators corresponding to any element of the algebra, $a \in \altc$,
\be
\label{hattedopdef}
\widehat{a} = \int dt\, e^{-i H t} X^{i t} a X^{-i t} e^{i H t} \projcont_t . 
\ee
For our discussion, the $t$ integration can be taken to be over $(-\infty, \infty)$. If we were to move away from the limit in section \ref{subseclimit} and go back to the precise setting, the $t$-integral would need to be discretized and, importantly, cut off at large positive and negative values of $t$. But these effects are not relevant for the current discussion.

As mentioned previously, a version of the construction \eqref{hattedopdef}  appeared in previous construction of relational observables \cite{Papadodimas:2015xma,Papadodimas:2015jra,Bahiru:2022oas,Bahiru:2023zlc} where the limits on the $t$-integration were discussed in more detail. In \cite{Papadodimas:2015xma,Papadodimas:2015jra} a similar construction was used to make certain operators in the interior of the eternal black-hole commute with the left Hamiltonian and transform under the action of the right Hamiltonian instead. (This is reviewed in Appendix \ref{seceternalinterior}.) In \cite{Bahiru:2022oas,Bahiru:2023zlc}, this construction was used to obtain bulk operators that perturbatively commute with the ADM Hamiltonian. The equation \eqref{hattedopdef} is closer in spirit to \cite{Papadodimas:2015xma,Papadodimas:2015jra}, since it alters the commutator of the operator with the Hamiltonian rather than making it vanish.

A little algebra shows that the procedure of deforming commutes with products so that
\be
\widehat{a b} = \widehat{a} \widehat{b}.
\ee

The deformed operators are close to the original operators in the sense that their action reduces to that of the original operator on a $0$-eigenstate of the time operator,
\be
\widehat{a} |0_{\Top }, b \rangle = a |0_{\Top },  b \rangle = |0_{\Top }, a b \rangle.
\ee
However, the action of the deformed operators on a state of the form $|f, a \rangle$ as given above differs from that of the original operators.  With a little algebra we find that 
\be
\widehat{a} |f, b \rangle = \int dt\,f(t) e^{-i H t} X^{i t} a X^{-i t} b |\zerovac  \rangle.
\ee

The key property of the deformed operators is given by studying 
\be
e^{i H s} \widehat{a} e^{-i H s} |f,b \rangle = \int dt\, f(t) e^{-i H t} X^{i (t+s)} a X^{-i (t+s)} b |\zerovac \rangle.
\ee
This tells us that on the deformed operators \eqref{hattedopdef}, time translation acts like conjugation with the operator $X$. More precisely, time-evolving a deformed operator by an amount $s$ is like conjugating the original operator with $X^{i s}$ and then deforming it. As an operator equation we have
\be
\label{deformedtransform}
e^{i H s} \widehat{a}  e^{-i H s} = \widehat {a_{s}},
\ee
with
\be
a_{s} \equiv X^{i s} a X^{-i s}.
\ee

Therefore, as advertised, the deformation \eqref{hattedopdef} can be viewed as dressing the operators to the observer rather than the boundary. Consequently,  time translations on these operators act as the chosen automorphism generated by $X$.

\subsubsection{Deformed operators as twirled operators}

It is illuminating to rewrite the construction \eqref{hattedopdef} in another form: deforming the original operators as above is equivalent to conjugation by an appropriate unitary. We also use the term ``twirled'' to refer to this conjugation .  This will help in making the relationship of the deformed operators with the crossed product explicit.

We define 
\be
V = e^{-i (H - \Qop) \Top } X^{i \Top }  = \int dt\,e^{-i (H - \Qop) t} X^{i t} \projcont_{t} .
\ee
Note that we have 
\be
[H - \Qop, \Top ] = 0,
\ee
and also
\be
[X, \Top ] = 0,
\ee 
and so there is no subtlety in studying the exponentiated operators above, since the operators in the exponent commute. 

The formula \eqref{hattedopdef} for the deformed operators can be written as 
\be
\label{hatasconj}
\widehat{a} = V  a V^{\dagger}.
\ee
To prove this we simply expand the factors of $V$ as above to find
\be
\widehat{a} = \int d t d t' e^{-i (H - \Qop) t'} \projcont_{t'} X^{i t'} a X^{-i t} e^{i (H - \Qop) t} \projcont_{t}.
\ee
We use 
\be
[\projcont_{t'}, H - \Qop] = 0; \qquad [\projcont_{t'}, a] = 0; \qquad [\projcont_{t'}, X^{i t}] = 0, \qquad [\Qop, X^{i t}] = 0,
\ee
and
\be
\projcont_{t'} \projcont_{t} = \delta(t-t') \projcont_t.
\ee
Doing the trivial integral over $t'$ now shows the equivalence of the formula \eqref{hatasconj} with \eqref{hattedopdef}.

The relation \eqref{deformedtransform} is easy to show in this notation. Using \eqref{qshift} and \eqref{hshift}, we see that
\be
V^{\dagger} e^{-i H t} =  e^{-i \Qop t} X^{-i t} V^{\dagger}; \qquad e^{i H t} V  =  V X^{i t} e^{i \Qop t}.
\ee
Using this, we simply commute the time-translation operators in \eqref{deformedtransform} past the factors of $V$ and $V^{\dagger}$ and use the fact that $\Qop$ commutes with elements of $\altc$ and $X$ to find \eqref{deformedtransform}.

\subsubsection{Result for the algebra}

We now define the algebra of the deformed operators
\be
\albarr = \{\widehat{a}: a \in \altc\}.
\ee
But we can now simply append the operators $e^{i H s}$ to this algebra 
\be
\label{fullcomplement}
\albarrfull = \text{span} \{\widehat{a} e^{i H s}: a \in \altc, s \in \mathbb{R}\}.
\ee

This algebra is closed under products since 
\be
\widehat{a} e^{i H s} \widehat{b} e^{i H r} = \widehat{a} \widehat{b_s} e^{i H (s + r)} = \widehat{a b_s} e^{i H (s + r)},
\ee
and $\widehat{a b_s} \in \albarr$. 

The algebra shown in \eqref{fullcomplement} is our final result for the algebra of the time band. Although the algebra is defined for boundary operators, physically, it is natural to associate this algebra in the bulk with the region $\band$. This completes the first part of our problem.  We now turn to the commutant of this algebra.

\subsection{Algebra in the bulk diamond}

We will identify the algebra in the bulk diamond, $\diam$, as the commutant of the algebra $\albarrfull$.

\subsubsection{Commuting operators}

We seek to list  operators that commute with all operators in $\albarrfull$. 

We start by examining the HKLL operators in the diamond, $\diam$ that were introduced in \eqref{freefullhkll} and \eqref{commutantqft}. These operators commute with all elements of $\altc$. 
\be
[\phihkll(t, r, \Omega), a] = 0, \quad \forall a \in \altc,
\ee
where it is understood that  bulk coordinates $(t, r, \Omega)$ are restricted to be in $\diam$.

However, these operators do not commute with the boundary Hamiltonian because we have
\be
e^{i H s} \phihkll(t, r, \Omega) e^{-i H s} = \phihkll(t+s, r, \Omega).
\ee
Therefore, these operators do not commute with elements of $\albarrfull$ either.

We can repeat a version of the trick above to deform the operators so as to make their commutator with the Hamiltonian vanish
\be
\label{almostcross}
\whphihkll(t, r, \Omega) = \int ds\,e^{-i H s} \phihkll(t, r, \Omega) e^{i H s} \projcont_s .
\ee
Note that the unitary operator $X^{i s}$ does not appear in the construction of $\whphihkll$.

With the help of some manipulations that are almost identical to those presented in the previous subsection, we may also write
\be
\whphihkll(t, r, \Omega) = W \phihkll(t,r, \Omega) W^{\dagger},
\ee
with
\be
W \equiv e^{-i (H-\Qop) \Top }.
\ee
The key property of $W$ is that
\be
\label{wcommuth}
e^{i H t} W  = W  e^{i \Qop t} \qquad  W^{\dagger} e^{-i H t} = e^{-i \Qop t} W^{\dagger}.
\ee

Consequently, the deformed HKLL operators commute with the Hamiltonian.
\be
e^{i H s} \whphihkll(t, r, \Omega) e^{-i H s}  = \whphihkll(t, r, \Omega).
\ee

We can then construct the algebra corresponding to all polynomials of the deformed HKLL operators in the diamond
\be
\alr = \text{span}\{\whphihkll(t_1, r_1, \Omega_1), \whphihkll(t_1, r_1, \Omega_1)  \whphihkll(t_2, r_2, \Omega_2) \ldots \},
\ee
with $t_i, r_i, \Omega_i$ restricted to the $\diam$. Referring to the polynomial algebra of all HKLL operators in the diamond defined in \eqref{commutantqft}, this algebra can also be written as
\be
\alr = \{W a W^{\dagger} : a \in \albarqft \}.
\ee

This does not exhaust the set of operators that commute with $\albarrfull$. One might hope to add the operators $e^{i \Qop s}$ which commute with elements of $\altc$ and also with $H$. But this operator does not commute with the operators $V$ and $V^{\dagger}$ that enter while defining  $\albarr$. Instead we need to consider
\be
\label{qqtwirled}
V e^{i \Qop s} V^{\dagger} = W e^{i \Qop s} X^{-i s} W^{\dagger}.
\ee
The expression on the left hand side of \eqref{qqtwirled} shows that this operators commutes with all elements of $\albarr$. The form given on the right hand side shows that it also commutes with $H$ using \eqref{wcommuth}.  

\subsubsection{Result for the commutant}
The analysis above leads us to conclude that the most general algebra of operators that commute with $\albarrfull$ is given by operators of the form
\be
\label{commutant}
\alrfull = \text{span}\{A W e^{i \Qop s} X^{-i s} W^{\dagger}, ~~ A \in \alr\,  ~ s \in \mathbb{R}\},
\ee
for all values of $s$. This algebra is physically identified with the algebra of the bulk diamond, $\diam$. 

Alternately, we can write this algebra in the form
\be
\label{commutantalternate}
\alrfull = \text{span}\{ W a e^{i \Qop s} X^{-i s} W^{\dagger},~~a \in \albarqft, ~ s \in \mathbb{R}  \}.
\ee
Since conjugation by $X$ is also an automorphism for $\albarqft$, in this latter form, it is obvious that the algebra $\alrfull$ is closed under products.
 
This algebra was defined as the commutant of the algebra for the time band, but it is natural to associate it in the bulk with the region $\diam$. We now have a construction  of algebras in both $\band$ and $\diam$.  In the section below, we show that this construction is unitarily equivalent to the construction of \cite{Jensen:2023yxy}.

\subsection{Unitary transformations and the crossed product}

We can  unitarily transform the algebras $\albarrfull$ and $\alrfull$ to obtain equivalent representations.  We do this by conjugating with $V^{\dagger}$ and refer to this operation as ``untwirling''.  Let us first untwirl the algebra $\albarrfull$.

\paragraph{Untwirled algebra in $\band$.}
Conjugating elements of $\albarr$ with $V^{\dagger}$ simply undoes the deforming operation that we implemented.
\be
V^{\dagger} \hat{a} V = a.
\ee
The action of this conjugation on the Hamiltonian is more interesting.
\be
X^{-i \Top } e^{i (H-\Qop) \Top } e^{i H t} e^{-i (H-\Qop) \Top } X^{i \Top }   = X^{-i \Top }  e^{i \Qop t} X^{i \Top } = e^{i \Qop t} X^{i t},
\ee
where we assume that $[\Qop, X] = 0$

Now we see that the algebra has turned precisely into a crossed product. Denoting the new algebra by $\albarprime$, it can be written as
\be
\label{albarprimeresult}
\albarprime = \text{span}\{a e^{i \Qop t} X^{i t}, ~~a \in \altc, ~~ t \in \mathbb{R} \}.
\ee

\paragraph{Untwirled algebra in $\diam$.}
Now let us transform the algebra $\alrfull$ using the same unitary transformation. We find
\be
V^{\dagger} \widetilde{\phi}(t, r, \Omega) V = X^{-i \Top } \phihkll(t, r, \Omega) X^{i \Top }.
\ee
Untwirling the remaining part of the algebra is trivial since it just undoes the conjugation by $V$ in \eqref{qqtwirled}. 

So the full untwirled algebra can be written as
\be
\label{alrprimeresult}
\alrprime = \text{span}\{ X^{-i \Top } a X^{i \Top } e^{i \Qop s},~~a \in \albarqft, \quad s \in \mathbb{R} \}.
\ee
This also has the form of a  crossed product.

In the analysis above, we have kept $X$ arbitrary. If we choose $X = \Delta^{-1}$, the crossed product above becomes the modular crossed product. Then the answers for the algebras $\albarprime$ and $\alrprime$ match precisely with the algebras for the time band and its complement identified in \cite{Jensen:2023yxy}. Equation \eqref{alrprimeresult} should be compared with Equation 5.3 in \cite{Jensen:2023yxy} and Equation \eqref{albarprimeresult} should be compared with Equation 5.2 there with the identifications $\Qop \rightarrow \hat{q}$ and $\Top \rightarrow -\hat{p}$.

\paragraph{Unitarily equivalent quantities.}
While the algebras $\albarprime$ and $\alrprime$ are unitarily equivalent to the algebra $\albarrfull$ and $\alrfull$ respectively, the correlators of elements of the algebra are altered in general states. However, in the state $|\zerovac \rangle$ the correlators of twirled operators are the same as correlators of untwirled operators. This is because
\be
V | \zerovac \rangle = V^{\dagger} |\zerovac \rangle = |\zerovac \rangle,
\ee
using the identities
\be
\Delta |\zerovac \rangle = |\zerovac \rangle; \qquad \Top |\zerovac \rangle = 0. 
\ee

Moreover, in the section below we will compute traces and entropies. These quantities are unchanged by the twirling operation that relates the two representations of the algebras.

\section{Traces and entropies}
\label{secentropy}

In this section, we will construct a trace on the algebras described above. This trace will take on a continuous set of a values in the limit described above, which shows that in this limit these algebras are of type II. The formulas that we describe in this section are very similar to the formulas provided in \cite{Witten:2021unn, Chandrasekaran:2022cip, Chandrasekaran:2022eqq, Jensen:2023yxy}. Our objective here is to translate the formulas that appear in those works to our setting.  Furthermore, we would like to show that the traces and entropies that have appeared in the discussion of gravitational subalgebras in the literature are only sensitive to the structure of the effective Hilbert space presented above. As such, they are insensitive to the UV properties of the theory or the microscopic density of states about the observer.

To be clear: the entropy in the effective Hilbert space is already an interesting quantity. It is equivalent to a framework where we study quantum field theory in curved spacetime and include the leading-order gravitational back-reaction classically. As such, this entropy has been used, for example, to prove a generalized second law for black holes \cite{Ali:2024jkx,Faulkner:2024gst,Suneeta:2024nxa}, and to provide a novel regulator for entropy differences in quantum field theory \cite{Kudler-Flam:2023hkl}. 

\subsection{Traces, modular operators and density matrices for quantum-mechanical subsystems}

We start by briefly reminding the reader of some properties of traces and density matrices in the case of ordinary quantum mechanical subalgebras that, in the language of von Neumann algebras, are called type I algebras. Say that we have a quantum system in a state $|\Omega \rangle$, and we are interested in a subalgebra of observables that we denote by $\alset$. 

Then the analogue of the little Hilbert space that we have discussed above is the space
\be
H_{\text{eff}} = \text{span}\{\al | \Omega \rangle \}, \qquad \al \in \alset.
\ee
We can choose an orthogonal basis of elements  for this little Hilbert space so that
\be
\langle \Omega | \al_i \al_j^{\dagger} | \Omega \rangle = \delta_{i j},
\ee
and also define a trace via
\be
\tr(\al_k) = \sum_i \langle \Omega | \al_i \al_k \al_i^{\dagger} | \Omega \rangle.
\ee
This trace is really a trace in the little Hilbert space. Thus, for instance, it does not know about the dimension of the larger space within which this system might be embedded.

The density matrix is defined as the  element $\rho \in \alset$ with the property that
\be
\tr(\rho \al_i) = \langle \Omega | \al_i |\Omega \rangle, \qquad \forall \al_i.
\ee
The density matrix is a positive operator since for any other positive operator, including any projector $P$, we have  
\be
\tr(\rho P) = \langle \Omega | P | \Omega \rangle \geq 0.
\ee

We assume that the density matrix for the state $|\Omega \rangle$ has no zero eigenvalues. Then the state
\be
|\hat{\Omega} \rangle = \rho^{-{1 \over 2}} |\Omega \rangle,
\ee
is well-defined. We now see that the expectation values of operators in this special state, $|\hat{\Omega} \rangle$ simply correspond to the trace since
\be
\langle \Omega |\rho^{-{1 \over 2}} a \rho^{-{1 \over 2}} |\Omega \rangle = \tr(\rho \rho^{-1 \over 2} a \rho^{-{1 \over 2}}) = \tr(a).
\ee
The state $|\hat{\Omega} \rangle$ is called a tracial state for the algebra. 

Consider a  state that can be obtained by exciting the tracial state with an element of the algebra,
\be
|b \rangle = b |\hat{\Omega} \rangle.
\ee
The density matrix for this state can be seen to be simply
\be
\label{densitynewstate}
\rho_{b} = b b^{\dagger}.
\ee
This is clear because, for any $a$,
\be
\langle b | a | b \rangle = \langle \hat{\Omega} | b^{\dagger} a b | \hat{\Omega} \rangle = \tr(b^{\dagger} a b) = \tr(b b^{\dagger} a).
\ee

The modular operator is defined, just as above, as the operator with the property that
\be
\langle \Omega | a \Delta_{\Omega} b |\Omega \rangle = \langle \Omega | b a |\Omega \rangle.
\ee
The modular operator itself is not an element of $\alset$. Instead, we can write\be
\label{modopfactor}
\Delta_{\Omega} = \rho  (\rho')^{-1}.
\ee
Here $\rho'$ is the density matrix for the commutant of $\alset$ that can be shown to have the same eigenvalues as $\rho$.

Even in the case of type I algebras, only some operators have a good trace. For instance, in the case of a simple harmonic oscillator, the position operator $x$ does not have a good trace; on the other hand, the projector on the vacuum state has a trace that is simply $1$. 

Subalgebras that correspond to finite regions in quantum field theory are generically of type III and, for such algebras, there is no good notion of a trace at all. One of the reasons for the recent interest in gravitational subalgebras is that type II algebras do allow for a trace, although the trace is not unique and has a rescaling ambiguity. This trace can be used to define an entropy for states up to a constant related to this rescaling ambiguity.

\subsection{Trace for the algebra}
In this subsection, we will define a trace for the algebras $\albarprime$ and $\albarrfull$. Our strategy is simple. We will identify the modular operator for the state $|\zerovac \rangle$ and factorize it into an element of the algebra and the commutant to obtain a density matrix. The formulas above then yield a tracial state.

Before proceeding, we need to fix the ambiguity in the algebras that we have constructed. We fix the automorphism to be 
\be
X = \Delta^{-1}.
\ee
As we explain in section \ref{sectrmodular}, this is effectively the only choice of $X$ that leads to a well-defined trace for the dressed algebra. For notational convenience we also write
\be
\Delta = e^{-\hmodpsi}.
\ee

We claim that this operator is also the modular operator for the full algebra $\albarprime$ in the state $|\zerovac \rangle$ 
\be
\label{fullmodisdelta}
\Delta_{\zerovac} = \Delta.
\ee
The defining property of the extended modular operator is that it reverse the order of operators in a correlation function,
meaning that for generic elements 
$e^{i(\Qop + h)s}a$ and $be^{i(\Qop+h)t}$
in $\albarprime$, the modular operator
$\Delta_{\zerovac}$ satisfies
\be
\label{checkmodular}
\langle \zerovac| b e^{i (\Qop + \hmodpsi) t} \Delta_{\zerovac} e^{i (\Qop + \hmodpsi) s} a | \zerovac \rangle = \langle \zerovac | e^{i (\Qop+\hmodpsi) s} a  b e^{i (\Qop+\hmodpsi) t} | \zerovac \rangle.
\ee
With a little algebra we see that
\be
\label{rhscheck}
\langle \zerovac | e^{i (\Qop+\hmodpsi) s} a  b e^{i (\Qop+\hmodpsi) t} | \zerovac \rangle  = \delta(s+t) \langle \smobsst | a b | \smobsst \rangle.
\ee
Substituting  \eqref{fullmodisdelta} on the left hand side we see
\be
\begin{split}
&\langle \zerovac| b e^{i (\Qop + \hmodpsi) t} \Delta e^{i (\Qop + \hmodpsi) s} a | \zerovac \rangle = \langle t_{\hat{T}}, \mathbbm{1} | b_{-t} \Delta a_{s} | -s_{\hat{T}}, \mathbbm{1} \rangle \\&= \delta(s + t) \langle \smobsst | a_{s} b_{s} | \smobsst \rangle = \delta(s+t) \langle \smobsst | a b |\smobsst \rangle,
\end{split}
\ee
which equals \eqref{rhscheck}. This proves \eqref{fullmodisdelta}.

But the full modular operator clearly factorizes as
\be
\label{factdelta}
\Delta_{\zerovac} = e^{-(\Qop + \hmodpsi)} e^{\Qop},
\ee
corresponding to an element of the algebra which we identify as the density matrix, and an element of the commutant that we identify as the inverse. The factorization \eqref{factdelta} is not unique since it is possible to multiply both the density matrices --- for the algebra and the commutant ---  by any $c$-number to obtain the same modular operator.

The relation \eqref{modopfactor} allows us to define the tracial state
\be
|\Omtr \rangle =  e^{{\Qop \over 2}} |\zerovac \rangle.
\ee
The ambiguity in the density matrix corresponds to a rescaling ambiguity in the tracial state. 

We can now define a trace for elements of the algebra $\albarprime$ by taking the expectation value 
in $|\Omtr\rangle$.  Operators of the form $a \widetilde{f}(h+\Qop)$
form a basis for $\albarprime$, and for these operators the trace simplifies.
\be
\tr(a \widetilde{f}(\hmodpsi + \Qop)) = \langle \Omtr | a \widetilde{f}(\hmodpsi + \Qop) | \Omtr \rangle =  \langle \smobsst | a | \smobsst \rangle \int dq\,e^{q} \widetilde{f}(q).
\ee
This trace is ambiguous up to a multiplicative factor due to the ambiguity in $\rho$ mentioned earlier.

The convergence of the integral over $q$ requires the Fourier transform of $\widetilde{f}(q)$ defined by the relation
\be
\widetilde{f}(q)= \int ds\,f(s) e^{-i q s}; \qquad f(s) = \int{d q \over 2 \pi} \widetilde{f}(q) e^{i q s} ,
\ee
to be analytic in the strip $\text{Im}(s) \in (-1, 0)$. We then have,
\be
\int dq\,e^{q} \widetilde{f}(q)  = f(-i).
\ee

The trace can be shown to be cyclic. We find,
\be
\label{trace1}
\begin{split}
\tr(a \widetilde{f}(\hmodpsi + \Qop) b \widetilde{g}(\hmodpsi + \Qop)) &= \int d s dt  f(s) g(t)  \langle \zerovac| e^{\Qop} a e^{-i (\hmodpsi + \Qop) s} b e^{-i (\hmodpsi + \Qop) t} | \zerovac \rangle \\
&= \int d s \langle \smobsst | a b_{-s} | \smobsst \rangle f(s) g(-i - s).
\end{split}
\ee
On the other hand, with the other ordering for the two operators,
\be
\label{trace2}
\begin{split}
\tr(b \widetilde{g}(\hmodpsi + \Qop) a \widetilde{f}(\hmodpsi + \Qop) ) &= \int d s d t f(s) g(t)  \langle \zerovac| e^{\Qop}  b e^{-i (\hmodpsi + \Qop) t} a e^{-i (\hmodpsi + \Qop) s} | \zerovac \rangle \\
&= \int d t  \langle \smobsst | b a_{-t} | \smobsst \rangle f(-i-t) g(t).
\end{split}
\ee
The two expressions look different but we can shift the contour of the $t$ integral in the complex plane until $t=-i-s$ where $s$ is real. We also have from the KMS condition
for modular flow and the relation $h|\Psi\rangle = 0$
that 
\be
\langle \smobsst | b a_{s+i} | \smobsst \rangle= \langle \smobsst | a b_{-s} | \smobsst \rangle,
\ee
These two substitutions convert the final expression in \eqref{trace2} into the final expression in \eqref{trace1}.

\paragraph{Trace for $\albarrfull$.}
A trace on an algebra is just a linear functional that is cyclic. The trace for the algebra $\albarprime$ immediately leads to a trace for the algebra  $\albarrfull$. Denoting the latter trace by $\widetilde{\tr}$, we set
\be
\widetilde{\tr}(V a f(\hmodpsi + \Qop) V^{\dagger}) \equiv \tr(a f(\hmodpsi + \Qop)).
\ee
This trace is clearly cyclic by virtue of the property of cyclicity that we proved for the trace on $\albarprime$. 
\subsection{Density matrices for normalizable states \label{subsecdensnorm}}
In the previous subsection, we identified the density matrix for the state $|\zerovac \rangle$. But this state is not normalizable.  Using the formula \eqref{densitynewstate}, we can find density matrices for normalizable states.

For instance, we have often referred to the state $|\fullst \rangle$ that can be written as
\be
|\fullst \rangle = \widetilde{f}(\Qop + h) e^{-{\Qop + h \over 2}} |\Omtr \rangle.
\ee
This immediately yields a density matrix for the $\albarprime$ algebra
\be
\rho_{\fullst} = |\widetilde{f}(\hmodpsi+\Qop)|^2 e^{-(\Qop+\hmodpsi)}.
\ee
A more-general state is of the form
\be
|g, c \rangle = c \widetilde{g}(\Qop + h) e^{-{\Qop + h \over 2}} |\Omtr \rangle,
\ee
and therefore its density matrix is 
\be
\rho_{g, c} = c |\widetilde{g}(\hmodpsi+\Qop)|^2 e^{-(\Qop+\hmodpsi)} c^{\dagger}.
\ee

It is clear that this method can be generalized to the case where we consider a linear combination of states above, including those that display entanglement between the $\altc$ and $\honedim$ sectors of the Hilbert space.

Density matrices for the algebra $\albarrfull$ are related to those of $\albarprime$ by unitary conjugation.
\be
\rho^{\albarrfull}_{\fullst} = V \rho_{\fullst} V^{\dagger}; \qquad \rho^{\albarrfull}_{g,c} = V \rho_{g,c} V^{\dagger}.
\ee

\subsection{Entropies}
Finally, we can use the formula for the trace and expressions for density matrices to obtain entropies. Since entropies are unaffected by unitary conjugation it does not matter whether we use the density matrix for the algebra $\albarprime$ or $\albarrfull$. 

For instance, the entropy for $|\fullst \rangle$ is simply
\be
\label{entstate}
S_{\fullst} = -\tr(\rho_{\fullst} \log \rho_{\fullst}) = -\int dQ\,|\widetilde{f}(Q)|^2 \log(|\widetilde{f}(Q)|^2 e^{-Q}) .
\ee

We would like to make two important points. First, as our discussion should make clear, this entropy should {\em not} be confused with the entropy of the state in the full UV-complete Hilbert space. It is an effective entropy relevant for effective Hilbert space that we have constructed. It is insensitive to the density of states about the observer in the full quantum-gravity theory.

Second, due to the ambiguity in defining the density matrix, this entropy is also ambiguous and can always be shifted by a constant. Therefore the absolute value of \eqref{entstate} is not very meaningful. However, once we have fixed the value of \eqref{entstate}, it makes sense to study the entropy of other states and compare their difference. This can be done using the expressions for the density matrix of arbitrary states derived in section \ref{subsecdensnorm}. 

Under further assumptions on the state of the bulk system, these entropy differences can sometimes be related to area differences. We refer the reader to \cite{Chandrasekaran:2022eqq, Chandrasekaran:2022cip, Jensen:2023yxy} for further discussion.

\section{Equivalence of trace and modular crossed product} \label{sectrmodular}

In the above construction, there was freedom in the definition of 
the algebras $\albarrfull$ and $\albarprime$ to choose the automorphism $X$ with which to 
twirl the elements of $\altc$.   Different choices of this 
automorphism lead to different algebras, each of which is a crossed 
product of the original undeformed type $\tthr_1$ subalgebra by the automorphism
generated by $X$.  
As originally derived
by Takesaki \cite{Takesaki1973} and explained in a number of recent works on gravitational algebras \cite{Witten:2021unn, Chandrasekaran:2022cip, Chandrasekaran:2022eqq, Jensen:2023yxy, Kudler-Flam2023, Faulkner:2024gst}, 
if $X$ is chosen to generate a modular
automorphism, the resulting algebra is type $\ttwo_\infty$.  
This 
observation leads to a well-defined renormalized trace 
for the dressed algebra, and with it a good notion of 
density matrix and entropy.  

In this section, we point out that the modular automorphism
is unique in this regard: it is the only automorphism
of a type $\tthr_1$ factor that yields an algebra with 
a trace upon constructing the crossed product
(this result is somewhat complementary to the 
recent work \cite{AliAhmad:2024eun} which determines 
when a crossed product by a group containing 
the modular automorphism results in a semifinite
von Neumann algebra).  Hence, one way to 
fix the choice in the operator $X$ in the construction
above is to demand that the resulting algebra possesses
a trace.  This then uniquely fixes $X$ up to inner automorphisms
generated by unitaries in the undeformed subalgebra.  These 
perturbations by inner automorphisms do not affect the 
overall type of the resulting crossed product 
\cite{Takesaki1973}\cite[Theorem X.1.7]{takesaki2003theory}.
Modular automorphisms are  
characterized by the existence of a state (or, more 
generally, a weight) for which the automorphism
satisfies a KMS condition \cite[Section VIII.1]{takesaki2003theory}.  Hence, the KMS condition 
for the undeformed subalgebra is in a sense dual to the 
tracial property of the dressed algebra.  

To be precise, we will prove the following theorem.
\begin{thm}
Let $\ahkll$ be a type $\tthr_1$ factor, and let $\agrav
= \mathbb{R}\ltimes_\alpha\ahkll$ be the crossed 
product of $\ahkll$ by a one-parameter group of automorphisms $\alpha_s$.  
If $\agrav$ is a semifinite factor (i.e.\ a factor possessing a faithful, semifinite,
normal trace),
then $\alpha_s$ must generate the modular automorphism
of a weight on $\alg$.  
\end{thm}
This theorem more or less follows from the general 
structure theorem for type $\tthr$ algebras and 
Takesaki duality 
\cite{Takesaki1973}\cite[Theorem XII.1.1, Theorem X.2.3]{takesaki2003theory}.  
We will give a 
direct argument for it below that also 
relates the weight for which $\alpha_t$ is the modular 
flow to the trace defined on $\agrav$.  

In this paper, we are interested in the case $\ahkll=\altc$. However, since the theorem is more general and applies to any $\tthr_1$ factor, we denote
this algebra simply by $\ahkll$. 

The idea of the proof is to make use of the time operator 
$\Top $ which arises in the construction of the dressed 
algebra.  The time operator appears as the conjugate
to the $\Qop$ operator in the crossed product description
derived above, and hence it can be characterized abstractly
as the generator of the dual automorphism $\hat{\alpha}_s$ 
that is guaranteed to exist for a crossed product
algebra.  Using this, 
we can construct a second crossed product algebra 
$\alg_1 = \mathbb{R}\ltimes_{\hat{\alpha}}\agrav $ which,
by Takesaki duality, is isomorphic to $
\mathcal{B}(L^2(\mathbb{R}))\otimes \ahkll$.  We exhibit this 
isomorphism explicitly, and then show that there 
is a natural weight defined on $\alg_1$ dual 
to the trace on $\agrav$ whose modular flow
maps the $\alg$ subalgebra into itself.  This relies 
on the observation that the time operator generates
an automorphism of $\agrav$ that rescales 
the trace, which we show separately.  We additionally derive
that this modular flow agrees with the original automorphism
$\alpha$ on $\alg$, and further show that it arises 
directly from a weight defined only on $\alg$.  This then
demonstrates the claim that $\alpha$ is the generator of a 
modular automorphism on $\alg$.  

We start with the representation of the $\agrav$ 
algebra acting on the tensor product Hilbert space $\honedim \otimes \cal{H}_\ahkll$, generated by the operators
\beq
\agrav = \left \langle e^{i\Top h} \msf{a} e^{-i\Top h},
\Qop\right\rangle,
\eeq
where $\Qop$ and $\Top $ are the conjugate position
and momentum operators acting on $L^2(\mathbb{R})$, and 
$\msf{a}$ is an arbitrary operator from $\ahkll$.   The automorphism
$\alpha$ is generated by $h$, and acts as $\alpha_t(\msf{a})
= e^{i t h}\msf{a}e^{-it h}$.  Since conjugation by $e^{is\Top }$
shifts $\Qop$ and leaves the operators $e^{i\Top h}
\msf{a} e^{-i\Top h}$ invariant, it is clear it generates 
an automorphism of $\agrav$.  Because $\Top $ is conjugate
to $\Qop$, $[\Top , \Qop] = i$, 
it serves as the generator of the 
dual automorphism to $\alpha_t$, and hence 
we denote this automorphism as 
$\hat\alpha_s(\hat{\msf a}) = e^{is\Top }\hat{\msf a}
e^{-is\Top }$ for $\hat{\msf{a}} \in
\agrav$.  

By assumption, $\agrav$ is a semifinite factor,
and hence possesses a 
trace $\tr$, unique up to an over all rescaling.\footnote{The technical statement
is that if a factor admits a faithful, semifinite, normal trace, the trace 
is unique up to rescaling \cite[Corollary V.2.32]{TakesakiI}.}  
Since $\hat{\alpha}_s$ generates an automorphism
of $\agrav$, the combined functional 
$\tr\circ \hat{\alpha}_s(\cdot) \equiv \tr(\alpha_s(\cdot))$ is also a trace, since
\beq
\tr\left(\hat\alpha_s(\hat{\msf{a}}\hat{\msf{b}})\right)
=
\tr\left(\hat\alpha_s(\hat{\msf{a}})\hat\alpha_s(\hat{\msf{b}})\right)
=
\tr\left(\hat\alpha_s(\hat{\msf{b}})\hat\alpha_s(\hat{\msf{a}})\right)
=
\tr\left(\hat\alpha_s(\hat{\msf{b}}\hat{\msf{a}})\right).
\eeq
By uniqueness of the trace, it must be that 
$\tr\circ \hat{\alpha}_s = e^{f(s)} \tr$,
for some function $f(s)$.  
Furthermore, since $\hat{\alpha}_s$ is a one-parameter
group of automorphisms satisfying $\hat{\alpha}_{s+t}
=\hat{\alpha}_s\circ\hat{\alpha}_t$, we straightforwardly
derive that $f(s+t) = f(s) + f(t)$.  Along
with the requirement that $f(0) = 0$, this uniquely
fixes $f(s) = cs$ for some constant $c$.  
The case $c = 0$ is not possible since then $\hat{\alpha}_s$ would preserve the trace,
and, as discussed below, this leads to the conclusion
that the original algebra $\ahkll$ possesses a 
trace, violating the assumption that $\ahkll$ is 
type $\tthr$.  Hence we will assume $c$ is nonzero,
and by possibly rescaling the generator $\Top $ we
can normalize the automorphism so that $c=1$.  
Hence, $\hat{\alpha}_s$ rescales the trace
according to 
\beq \label{eqn:trscale}
\tr \circ \hat\alpha_s = e^{s} \tr.  
\eeq

We can realize the trace on $\wh\alg$ as the expectation
value in an unnormalized state $|\Omtr\rangle$,
\beq \label{eqn:trunnorm}
\tr(\msf{\hat{a}}) = \langle\Omtr| \msf{\hat{a}}|
\Omtr\rangle.
\eeq
This state is unnormalized since $\langle\Omtr|\Omtr
\rangle = \tr(\mathbbm{1})$, and the identity must 
have infinite trace due to the existence of the trace
scaling automorphism: $\tr(\mathbbm{1}) = 
\tr(\alpha_s(\mathbbm{1})) = e^s \tr(\mathbbm{1})$,
and this is only consistent if $\tr(\mathbbm{1}) = \infty$.  
The relations (\ref{eqn:trscale}) and 
(\ref{eqn:trunnorm}) imply that $\Top $ acts on 
$|\Omtr\rangle$ as $e^{-
is\Top }|\Omtr\rangle = e^{\frac{s}{2}}
|\Omtr\rangle$.  This suggests that 
$|\Omtr\rangle$ should take the form
$| e^\frac{Q}{2}, \Omhkll\rangle$, where $|\Omhkll\rangle$
is a (possibly unnormalized) state in $\cal{H}_\ahkll$
that defines a weight 
for the operators $\msf{a} \in \ahkll$, and 
$|e^{\frac{Q}{2}}\rangle$ is a unnormalized
wavefunction associated with 
$L^2(\mathbb{R})$ in the $\Qop$ basis. 
We would like to identify 
the state $|\Omhkll\rangle$ in an unambiguous 
way, and further show that $h$ generates the 
modular flow of $|\Omhkll\rangle$ as a weight
on $\alg$.  

To do this, we construct a second crossed 
product with respect to the $\hat{\alpha}_s$
automorphism on $\wh{\alg}$.  This is done by 
extending the Hilbert space with an additional
factor of $L^2(\mathbb{R})$, and introducing
position and momentum operators $\hat{y}$, $\hat{k}$
acting on this Hilbert space satisfying $[\hat{y},
\hat{k}] = i$.  The crossed product 
$\alg_1 := \mathbb{R}\ltimes_{\hat{\alpha}}\wh{\alg}$
is then generated by the operators,\footnote{Equivalently,
these are all operators of the form
$e^{i\hat{k}\Top }\hat{\msf{a}}e^{-i\hat{k}\Top }$
with $\hat{\msf a}\in \wh{\alg}$, along with functions
of $\hat{y}$. These operators are equivalent 
to those displayed in (\ref{eqn:A1ops}) because
$\Top $ commutes with all operators 
of the form $e^{i\Top h}\msf{a}e^{-i\Top h}$,
$\msf{a}\in\alg$.}
\beq \label{eqn:A1ops}
\alg_1 = \left\langle e^{i\Top h} \msf{a}
e^{-i\Top h}, \Qop-\hat{k}, \hat{y}\right\rangle,
\eeq
acting on the Hilbert space $\hs_1 
=L^2(\mathbb{R}_y)\otimes L^2(\mathbb{R}_Q) \otimes \cal{H}_\ahkll$.
There is a natural
weight defined on the algebra $\alg_1$ arising 
as the dual weight of $\tr$ on the $\wh{\alg}$
subalgebra; its unnormalized vector representative 
$|\Omone\rangle$
is given  by 
\beq \label{eqn:Om1}
|\Omone\rangle = |0_k\rangle\otimes |\Omtr\rangle,
\eeq
where $|0_k\rangle$ is the delta-function
normalized zero momentum eigenstate 
in $L^2(\mathbb{R}_y)$.  We would like to compute
the modular operator for this weight.  This can 
be done by determining how the 
Tomita operator\footnote{Recall
that the Tomita operator is an antilinear 
operator associated with a cyclic-separating 
vector $|\Phi\rangle$ defined by the 
relation $S_\Phi \msf{a}|\Phi\rangle = \msf{a}^\dagger
|\Phi\rangle$ for all operators $\msf{a}$ in a 
von Neumann algebra $\alg$.  Its polar decomposition
gives $S_\Phi = J_\Phi \Delta_\Phi^{\frac12}$,
where $J_\Phi$ is the antiunitary 
modular conjugation, and $\Delta_\Phi = S_\Phi^\dagger
S_\Phi$ is the modular operator.  See \cite{Witten:2018zxz} for 
a more thorough introduction to Tomita-Takesaki theory.} $S_1$ acts on operators of the form
$\msf{a}_1 = e^{i\hat{k}\Top }\hat{\msf{a}}
e^{-i\Top \hat{k}} e^{i u\hat{y}}$. Acting
on $|\Omone\rangle$, this gives
\beq\label{eqn:a1Om1}
\msf{a}_1|\Omone\rangle
= e^{i \hat{k} \Top }\hat{\msf a}e^{-i\hat{k}\Top }
e^{i u\hat{y}}|0_k, \Omtr\rangle
= e^{i \hat{k} \Top }\hat{\msf a}e^{-i\hat{k}\Top }|u_k, \Omtr\rangle
=
e^{i \hat{k} \Top }\hat{\msf a}e^{\frac{\hat{k}}{2}}|u_k,\Omtr\rangle,
\eeq
while $\msf{a}_1^\dagger$ gives
\beq
\msf{a}_1^\dagger |\Omone\rangle
= e^{-i u\hat{y}} e^{i\hat{k} \Top }\hat{\msf{a}}^\dagger
e^{-i\hat{k}\Top }|0_k, \Omtr\rangle
=
e^{-i u\hat{y}} e^{i\hat{k} \Top }\wh{J}\hat{\msf{a}}|0_k, \Omtr\rangle
=
\wh{J} \hat{\msf{a}}|u_k, \Omtr\rangle,
\eeq
where in the second equality we made use of the fact 
that $|\Omtr\rangle$ is tracial, 
so $\Delta_{\Omtr} = \mathbbm{1}$ and 
$S_{\Omtr} = J_{\Omtr}$.  
Since the Tomita operator for $\alg_1$ 
satisfies $S_1 \msf{a}_1|\Omone\rangle
 = \msf{a}_1^\dagger|\Omone\rangle$, these 
 equations determine $S_1$ and the modular 
 operator $\Delta_1$ to be
 \beq \label{eqn:S1}
S_1 = \wh{J}e^{-i\hat{k}\Top } e^{-\frac{\hat{k}}{2}},
\qquad \Delta_1 = S_1^\dagger S_1 = 
e^{-\hat{k}}.
 \eeq
This shows that $\hat{k}$ is the generator
of modular flow for the weight associated with $|\Omone\rangle$.\footnote{Alternatively,
we can derive this expression for the modular operator by invoking Takesaki's dual 
weight theorem \cite[Theorem X.1.17]{takesaki2003theory}, which states that modular
flow of the dual weight reproduces modular flow of $|\Omtr\rangle$ on the $\wh{\alg}$
subalgebra, which is trivial since $|\Omtr\rangle$ is tracial, and acts on
the operators $e^{-iu\hat{y}}$ according to $\Delta_1^{is} e^{-iu\hat{y}} \Delta_1^{-is} = 
(D (\tr\circ \hat{\alpha}_u): D\tr)_s e^{-iu\hat{y}}$, where $(D (\tr\circ \hat{\alpha}_u): D\tr)_s
=e^{ius}$ is the Connes cocycle derivative between the two weights
$\tr\circ\hat{\alpha}_u$ and $\tr$ \cite{Connes1973}\cite[Section
VIII.3]{takesaki2003theory}.  
}

We then want to determine how this modular flow
restricts to the algebra $\alg$ in the 
tensor product decomposition $\alg_1 \equiv \mathcal{B}(L^2(\mathbb{R})\otimes 
\alg$.
To exhibit this isomorphism, we rewrite the 
operators generating $\alg_1$ as 
\beq
\alg_1 = \left\langle e^{i(\Top +\hat{y})h}\msf{a}
e^{-i(\Top + \hat{y}) h}, 
\Qop - \hat{k}, \hat{y}\right\rangle.
\eeq
Since $\Top + \hat{y}$ commutes with 
both $\Qop-\hat{k}$ and $\hat{y}$, the operators 
$e^{i(\Top +\hat{y})h}\msf{a}
e^{-i(\Top + \hat{y}) h}$ generate an algebra $\tilde\alg$
isomorphic to $\alg$ and commute with the algebra 
generated by $\Qop-\hat{k}$ and $\hat{y}$. 
Since these latter two operators satisfy 
a canonical commutation relation $[\Qop-\hat{k},\hat{y}]
= i$, the algebra they generate is isomorphic to the 
type $\tone_\infty$ algebra $\mathcal{B}(L^2(\mathbb{R}))$.  
Modular flow with respect to $\hat{k}$ then acts 
on an operator in $\tilde \alg$ as 
\beq
e^{is\hat{k}} e^{i(\Top +\hat{y})h}\msf{a}
e^{-i(\Top + \hat{y}) h} e^{-is{\hat{k}}}
=
e^{ish}e^{i(\Top +\hat{y})h}\msf{a}
e^{-i(\Top + \hat{y}) h}e^{-ish},
\eeq
showing that it not only preserves the algebra $\tilde \alg$,
but also generates the same transformation as $h$ acting 
on the original algebra.  This is nearly enough to 
conclude that $h$ generates a modular automorphism of a weight
on $\alg$; we need only exhibit the weight.

Intuitively, the vector $|\Omone\rangle$ should have the 
form
\beq
|\Omone\rangle = |0_k, e^{\frac{Q}{2}}, \Omhkll\rangle,
\eeq
where $|\Omhkll\rangle$ is the vector representation
of the weight on $\alg$ that we are interested in.  However, 
we cannot restrict $|\Omone\rangle$ to $\alg$ directly since it is 
not semifinite on this subalgebra; it picks up the divergent 
norm of the state $|\chi\rangle =|0_k,e^{\frac{Q}{2}}\rangle$ whenever it is 
evaluated on operators in $\alg$.  Hence, to obtain
the state $|\Omhkll\rangle$, we would like to divide out by this
divergent norm.  

The formal argument for how to do this is 
to construct an operator-valued weight\footnote{Operator-valued
weights capture the idea of tracing out 
part of the algebra to arrive at 
operators in a subalgebra.  Formally, it is 
defined as a map from a dense subalgebra 
of $\alg_1$ to $\tilde\alg^c$ satisfying 
the bimodule property: $T(\msf{bac}) = \msf{b}T(\msf{a})
\msf{c}$ whenever $\msf{b}, \msf{c} \in 
\tilde{\alg}^c$.  See 
\cite[Section IX.4]{takesaki2003theory}} $T$ from $\alg_1$ to 
the relative commutant subalgebra $\tilde{\alg}^c = \tilde{\alg}'
\wedge \alg_1 = \left\langle \Qop-\hat{k}, \hat{y}\right\rangle$,
and show that the weight $\omega_1 = \langle\Omone|\cdot|\Omone
\rangle$ satisfies $\omega_1 = \omega_\chi\circ T$, where $\omega_\chi$ is a 
weight on $\tilde{\alg}^c$.  This weight can be explicitly identified;
it is simply given by $\omega_\chi = \langle 0_k, e^{\frac{Q}{2}}|
\cdot |0_k, e^{\frac{Q}{2}}\rangle =\langle\chi|\cdot|\chi\rangle$.  
To check that this is correct,
we can compute the Tomita operator $S_\chi$ for this: taking 
a generic operator $\msf{c} = e^{-iu(\Qop-\hat{k})} e^{iv\hat{y}}$
in $\tilde{\alg}^c$, we have 
\begin{align}
\msf{c}|\chi\rangle
&=
e^{-iu(\Qop-\hat{k})} e^{\frac{\Qop}{2}}|v_k, 0_{\Top}\rangle
=
e^{\frac{\Qop}{2}} e^{iuv}|v_k, u_{\Top}\rangle
=
e^{\frac{\Qop}{2}} e^{i\Top \hat{k}}|v_k, u_{\Top}\rangle
=
e^{\frac{\hat{k}}{2}}e^{i\Top \hat{k}} e^{\frac{\Qop}{2}}|v_k, u_p\rangle
\\
\msf{c}^\dagger|\chi\rangle
&=
e^{-iv\hat{y}} e^{\frac{\Qop}{2}}|0_k, -u_{\Top}\rangle
=
e^{\frac{\Qop}{2}}|-v_k, -u_{\Top}\rangle
=
J e^{\frac{\Qop}{2}}|v_k, u_{\Top}\rangle,
\end{align}
where $J$ is the antiunitary time-reversal operator that acts
as $J|u_{\Top},
v_k\rangle = |-u_{\Top}, -v_k\rangle$.  Since the Tomita
operator acts as $S_\chi \msf{c}|\chi\rangle = \msf{c}^\dagger
|\chi\rangle$, these relations determine it and the modular 
operator to be
\beq
S_\chi = Je^{-i\Top \hat{k}}e^{-\frac{\hat{k}}{2}}
\qquad \Delta_\chi = S_\chi^\dagger S_\chi = e^{-\hat{k}}.
\eeq
This modular operator agrees with $\Delta_1$ from (\ref{eqn:S1}),
and hence generates the same flow on $\tilde{\alg}^c$.  This 
is precisely the condition needed to apply
Haagerup's theorem \cite{Haagerup1979II}\cite[Theorem IX.4.18]{takesaki2003theory}
for the existence of an operator-valued 
weight $T$ from $\alg_1$ to $\tilde{\alg}^c$ satisfying
\beq
\omega_1 = \omega_\chi\circ T.
\eeq

Since $\tilde{\alg}^c$ is a factor, the restriction of $T$ to 
$\tilde{\alg}$ will map each operator to something proportional
to the identity; hence, $T$ defines a weight on $\tilde\alg$.  
The modular flow of this weight agrees with that of $|\Omone\rangle$,
and hence corresponds to the flow generated by $h$.  We 
see then that the operator-valued weight $T$ corresponds to the 
idea of dividing out by the divergent norm
of the state $|\chi\rangle$.  This 
is precisely the weight on $\alg$ we were seeking, and hence 
have shown that $h$ is the generator of a modular flow.  
This concludes the argument that the only crossed 
product of a type $\tthr_1$ factor that results in an 
algebra with a trace is the crossed product by a modular flow.  

Finally, to complete the argument, we need to rule out the 
case where the dual automorphism $\hat{\alpha}_s$ preserves 
the trace on $\wh{\alg}$.  In this case, the generator of the 
dual automorphism $e^{-is\Top }$ preserves the unnormalized 
tracial weight $|\Omtr\rangle$.  Hence, we can follow
the same steps as before in constructing the second 
crossed product algebra $\alg_1$ as in (\ref{eqn:A1ops})
and the dual weight $|\Omone\rangle$ as in (\ref{eqn:Om1});
however, when computing the Tomita operator, no
factor of $e^{-\frac{\hat{k}}{2}}$ appears in equation (\ref{eqn:a1Om1}).
Instead, we simply find
\beq
S_1 = \wh{J} e^{-i\hat{k}\Top }, \quad \Delta_1 = \mathbbm{1}.
\eeq
We see that the modular operator is trivial, meaning that the 
weight $|\Omone\rangle$ defines a  trace on the algebra
$\alg_1$.  By Takesaki duality, $\alg_1 = 
\mathcal{B}(L^2(\mathbb{R}))\otimes\alg$, and since 
$\mathcal{B}(L^2(\mathbb{R}))$ is type $\tone_\infty$, we 
see that $\alg$ cannot be type $\tthr$: tensoring a type 
$\tthr$ factor with a type $\tone_\infty$ factor results in
an isomorphic algebra, 
but $\alg_1$ cannot be type $\tthr$ since we have shown
$\alg_1$ possesses a trace.
Hence, we see that the assumption that the dual automorphism
$\hat{\alpha}_s$ is trace-preserving is inconsistent
with $\alg$ being type $\tthr$; thus, it must be that 
$\hat{\alpha}_s$ does not preserve the trace.  

\section{Higher orders in perturbation theory and nonperturbative effects} \label{sechigherorder}

It is commonly believed that the construction of the type II algebra presented above can be extended to all orders in perturbation theory. A weaker assertion is that even if the structure of the algebra is hard to parse, it should be possible to systematically understand the higher-order corrections to the entropy of a state. In this section we would like to clearly lay out some technical challenges that must be surmounted to extend this construction to higher orders in perturbation theory. To be clear: we do not see any of these challenges as insuperable obstacles and the discussion below does not invalidate the idea that clever technical advances will help in systematically understanding all higher-order perturbative effects.

Many of the issues below relate to the emergence of an effective time operator in our setup. This requires the factorization of the limiting Hilbert space displayed in \eqref{tensordecomp} and also other properties of the algebra. Naively, this structure does not appear to persist beyond lowest orders in perturbation theory. 

For previous ideas  on going beyond the leading large $N$ limit, see \cite{Witten:2023xze, Faulkner:2022ada}.

\paragraph{Loss of separation between Hamiltonian and light operators.}
The construction presented in section \ref{secobserver} started with a set of light operators, $\altc$, which explicitly excluded the Hamiltonian. However, beyond leading order in perturbation theory, there is no clear separation between light operators and the Hamiltonian. 

From the bulk perspective, the Hamiltonian is the integral of a component of the asymptotic metric. Therefore, separating the Hamiltonian from other operators requires us to separate the metric from other degrees of freedom. But the metric couples universally to matter in a gravitational theory. So, the interaction of matter fields can always produce a metric fluctuation. 

This issue can be seen clearly in the dual CFT. Consider two light operators at positions $x$ and $0$ on the boundary in projective coordinates. Then we have the following term in the  OPE expansion on the boundary
\be
\label{opeexpansion}
O(x) O(0) = \ldots +  {1 \over c |x|^{2 \Delta - d}} {x^{\mu} x^{\nu} \over |x|^2}  T_{\mu \nu}(0) + \ldots,
\ee
where $c$ is a constant proportional to the central charge. 
Integrating the stress tensor over a boundary Cauchy slice produces the Hamiltonian and so the product of two light operators already implicitly contains the Hamiltonian at higher orders in perturbation theory. 

This problem can also be understood through simple physical reasoning.  In section \ref{secobserver}, we introduced the operator $\Qop$ that played a key role in our construction. This operator can be thought of as a coarse-grained version of the boundary Hamiltonian that is sensitive to the energy of the background observer but insensitive to low-energy excitations on top of the observer's state. At leading order in perturbation theory, this is acceptable since the two energies are parametrically separated. But this separation is nontrivial at subleading order.

\paragraph{Overlaps at long times.}
Relatedly, the direct product structure of the effective Hilbert space relied on the observation that the time-translated little Hilbert spaces rapidly become orthogonal. 
\be
\langle \obsst | e^{-i H t} a | \obsst \rangle \ll 1, \qquad \text{if}~~ t \gg {1 \over \obsenfluct},
\ee
when $a$ is a light operator separate from the Hamiltonian. 
But this is not true if $a$ itself can contain powers of the Hamiltonian. For instance, 
\be
\langle \obsst | {e^{-i H t} \over H} | \obsst \rangle =  \langle \obsst |\left( -i \int_0^{t}ds\, e^{-i H s}  + {1 \over H} \right)| \obsst \rangle = \Or[1 \over \delta E],
\ee
which is nontrivial when we go beyond leading order in perturbation theory. The estimate is made by noting that $\langle \obsst | e^{-i H s} | \obsst \rangle$ is close to $1$ for $s \in (0, {1 \over \delta E})$. The nontrivial overlap above is present even if $t$ is large. 

\paragraph{Interactions between light operators and the observer.}
The direct product structure relied on our demonstration that the space \eqref{htdef} was dense in the space \eqref{htotherorderdef} i.e. the set of states produced by first acting with time translation and then acting with a polynomial of simple operators was effectively the same as the set of states produced by performing these operations in the reverse order. This required us to make some assumptions about the matrix elements of the set of simple operators in the observer state. Physically, these assumptions were justified by assuming that the observer's energy interacts only weakly with the insertion of other probes. However, this might not be true once we include gravitational corrections at next nontrivial order in perturbation theory. 

\paragraph{Discreteness of time.}
We expect the observer's clock to be accurate up to time intervals of size $\Or[{1 \over \delta E}]$. Mathematically, this is related to the issue that
\be
\langle \obsst | e^{-i H t} | \obsst \rangle \approx  1, \qquad \text{if}~~ t \ll {1 \over \obsenfluct}.
\ee
Therefore the little Hilbert space at a given time cannot be meaningfully separated from the little Hilbert space at a neighbouring time separated by an interval less than $\Or[{1 \over \obsenfluct}]$. 

At leading order in perturbation theory, this issue can be neglected. The limit taken in section \ref{subseclimit} drops effects of size $\Or[{1 \over \obsenfluct}]$, which allows us to imagine that the observer's time is continuous. But at higher orders in perturbation theory we cannot drop these effects. Therefore, it is not clear how to extend the notion of a time operator at higher orders in perturbation theory.

\paragraph{Edge effects.}
They key problem in defining an algebra for the time band and its commutant is to find operators that have a specified commutator with the Hamiltonian. In this paper, we defined such operators by exploiting the direct product structure of the effective Hilbert space and introducing a time operator and its conjugate. Even if this structure is lost, one might hope to use the following robust construction to define a commutant. 

Let us add the Hamiltonian and cutoff polynomials that involve the Hamiltonian to the set $\altc$ and call this new set $\altc'$. We expect that as long as $\dim(\altc') \ll e^{S}$, the state $|\obsst \rangle$ will remain separating in that
\be
\label{stillseparating}
a'   |\obsst \rangle \neq 0 \quad \forall a' \in \altc'.
\ee

This property allows us to define a commutant for $\altc'$. For instance, say that we seek an operator that commutes with all elements of $\altc'$ and maps $|\obsst \rangle$ to some specified state $|\obsst_a \rangle$. Then we can simply define such an operator by its action on the Hilbert space through
\be
\widetilde{a}  b' |\obsst \rangle =   b' |\obsst_{a} \rangle.
\ee
for any $b' \in \altc'$. This provides a consistent set of linear equations for the operator $\widetilde{a}$ by virtue of \eqref{stillseparating}. Moreover, these relations completely specify the  $\widetilde{a}$ within the effective Hilbert space, since they specify its action on all elements of this space.  

This is reminiscent of the mirror operator construction of \cite{Papadodimas:2013wnh,Papadodimas:2013jku}. It is not as elegant as simply twirling the original operator with a unitary that involves a time operator. But the virtue of this construction is that it remains valid even if the definition of the time operator suffers from the subtleties discussed above.

But $\altc'$ is not an algebra because the cutoffs that we place on the set prevent it from being closed under products. In particular if $a,b \in \altc'$, we might still have $a b \notin \altc'$. We cannot simply get rid of the cutoff since we might then lose \eqref{stillseparating}.  Therefore the equations above only make sense provided we ignore edge effects. 

In this paper, and in all the extant literature so far, such edge effects have been ignored. It is clearly justified to do so at leading order in perturbation theory. It is not clear whether these edge effects can be neglected at all orders in perturbation theory.

Moreover, a careful treatment of edge effects should also make physical sense.  For instance, it is technically convenient to work with polynomials of light operators where a cutoff can be placed on the order of the polynomial. However, physical measurements might be more easily described in terms of projective measurements.  So it is necessary to understand how one should carefully restrict the set of observables at finite $N$ and then take a limit as $N \rightarrow \infty$ so that the results obtained are valid to all orders in the ${1 \over N}$ expansion. 

\subsection{Nonperturbative effects}

Nonperturbatively, it is clear that the algebra of the time band is the full algebra of the theory. This follows from AdS/CFT since any Cauchy slice of the boundary contains all operators. However, this property is not mysterious even from the point of view of the bulk gravitational theory. We quickly review, following \cite{Laddha:2020kvp}, how this result follows from a few reasonable physical assumptions about the UV-complete theory without invoking AdS/CFT. 

First, we assume that it makes sense to discuss the asymptotic algebra of observables in the UV-complete theory i.e. it is possible to define good asymptotic observables by extrapolating bulk observables using \eqref{extrapolatedict}.  This reflects the point of view that, even in a theory of quantum gravity, it makes sense to fix asymptotic boundary conditions and assume that asymptotically we have an AdS spacetime. 
We define the algebra of the time band via smeared boundary operators
\be
\alset = \text{span}\{O(f_1), O(f_1) O(f_2),  \ldots, O(f_1) O(f_2) \ldots O(f_n) \ldots \},
\ee
where $f_i$ have support in the time band $\bandset$.  Since we are now interested in the fine-grained algebra, we place {\em no} restriction on $n$ and we do {\em not} remove the Hamiltonian from the list of operators that enter $\alset$. 

Second, we assume that the energy is bounded below and that the vacuum is unique. This is a reasonable assumption about any sensible UV-completion, since the low-energy theory in AdS is clearly gapped.  One may now consider the Hilbert space formed by acting with all boundary operators on this vacuum that we denote by $|\Omega \rangle$.
\be
{\cal H} = \text{span}\{O(g_1), O(g_1) O(g_2),  \ldots, O(g_1) O(g_2) \ldots O(g_n) \ldots \} | \Omega \rangle.
\ee
In defining this Hilbert space, not only do we not place a cutoff on $n$, we allow the functions $g_i$ to have support on the entire boundary and not just on the time band. Consequently, it is clear that this Hilbert space constitutes a superselection sector since time evolution only takes a smearing function to another valid smearing function.

Now the positivity of energy implies that if we consider the space
\be
{\cal H}' = \alset | \Omega \rangle,
\ee
then ${\cal H}'$ is dense in ${\cal H}$. (See the Appendix of \cite{Laddha:2020kvp} for a proof.) This means that given any state $|n \rangle \in {\cal H}$ we can find $X_n \in \alset$ so that 
\be
X_n | \Omega  \rangle \doteq | n \rangle,
\ee
where $\doteq$ means that equality holds to any desired precision.

So far we have not invoked gravitational physics in the bulk, which we now do. In a theory of gravity the Hamiltonian itself is an element of the algebra of the time band.  This is a manifestation of the bulk Gauss law. Although this is a property of the UV theory, one needs to assume that even in the full UV-complete theory at least the projector on the vacuum remains an element of $\alset$.\footnote{In fact, it is sufficient if the projector onto any state with strictly finite energy be an element of $\alset$.} 
\be
\label{projvacinalset}
\projvac = |\Omega \rangle \langle \Omega | \in \alset.
\ee
By the Born rule, $\projvac$ simply tells us the probability that a measurement of the energy will yield the vacuum energy. Therefore, it makes sense from a physical perspective, it makes sense to include this operator in any complete description of observables. From an algebraic perspective, $P_0$ is a minimal projection,  and hence including it in our algebra 
implies that the algebra is type $\tone$, which is consistent with the expectation that the nonperturbative global quantum gravity algebras are type $\tone$. 

We now find that any operator in the Hilbert space $|n \rangle \langle m |$ can be represented as
\be
| n \rangle \langle m | \doteq X_n \projvac X_m^{\dagger}.
\ee

This result is the precise statement of holography of information in an asymptotically AdS spacetime. It tells us that all observables in the bulk have a holographic description and therefore all bulk information is available at the boundary. However, it does not give us a dynamical boundary dual for the bulk theory. 

Therefore, at the nonperturbative level, rather robust properties of the bulk gravitational theory tell us that the algebra of the time band has no commutant. Nonperturbatively, it only makes sense to study subalgebras associated with subregions of the boundary. For instance, if we consider a spatially bounded subregion of the boundary then the $\projvac$ is not an element of its algebra. So the argument above breaks down and this algebra has a commutant even nonperturbatively.

\section{Discussion}
\label{secdiscussion}

In this paper we framed and resolved a puzzle related to the algebra of a boundary time band in a gravitational theory in asymptotically anti-de Sitter space. This algebra can be associated with the algebra of an annular
region in the bulk that is the complement of a bulk causal diamond. Bulk
gravitational algebras of 
this form have been studied previously \cite{Jensen:2023yxy}. At first sight, they appear to suffer from an inconsistency: in a theory of gravity, the ADM Hamiltonian is an observable in the boundary time band, and its action can move an element from inside the time band to outside. This puzzle does not require us to postulate the existence of a holographic dual
but is implicit in the properties of gravity as has been discussed extensively in the recent literature on holography of information \cite{Laddha:2020kvp}.

The key physical observation that helps to resolve this puzzle is that the holographic properties of gravity are important in some regimes but can be obscured in others. They are important when we study nonperturbative questions, such as those related to information loss in black-hole evaporation \cite{Raju:2021lwh,Raju:2020smc},  and they are also important when we study perturbative excitations about the AdS vacuum \cite{Chowdhury:2020hse,Chowdhury:2022wcv,Gaddam:2024mqm}. On the other hand, if we study perturbative excitations about a heavy background state and coarse grain the boundary algebra, the holographic properties of gravity can be neglected \cite{Bahiru:2022oas,Bahiru:2023zlc}. In this respect, holography in gravity is somewhat like unitarity in ordinary quantum field theories: unitarity is important for nonperturbative questions, and it can also be verified explicitly in perturbation theory. But in the presence of a heavy background, when we coarse-grain the set of observables, physics is effectively dissipative. 

The heavy background state we study is like the observer that has appeared previously in the literature \cite{Chandrasekaran:2022cip}. Our approach is somewhat novel since we do not introduce the observer through an external particle, or by postulating auxiliary decoupled degrees of freedom \cite{Chataignier:2024eil,DeVuyst:2024pop,DeVuyst:2024grw,Kaplan:2024xyk}. Instead we study a state within the theory (as was also done in a cosmological context in \cite{Chen:2024rpx}) and examine the conditions that must be placed on such a state for it to serve as a good observer. 

If the observer's state has a sufficiently large spread in energies, then time-translated versions of this state are almost orthogonal to the original state.  This is the key property that allows us to define a time operator, $\Top $.  The conjugate to this time operator, $\Qop$ can be thought of a coarse-grained Hamiltonian that is sensitive to the energy of the observer but not to the energy of low-energy excitations about the observer. 

By deforming light boundary operators with an appropriate unitary operator built from $H, \Qop, \Top$ and the generator of an automorphism $X$, we obtained a new set of operators for which the action of the Hamiltonian acts as an automorphism. This allowed us to define a closed algebra for the time band and also allowed us to obtain a nontrivial commutant. 

Our construction allows us to choose the automorphism that is generated on deformed operators by the boundary Hamiltonian. When we choose this automorphism to be the modular automorphism, we find that our algebras are unitarily equivalent to those identified in \cite{Jensen:2023yxy}. 

We justified our choice of dressing by demanding the resulting algebra possesses  a trace, and showed that this uniquely determines the dressing so that time translation acts as a modular automorphism on the algebra. It would be preferable to have 
a more concrete criterion for selecting this dressing, which is a question we leave to future work. 

Our analysis establishes a link between the gravitational crossed product that has been studied extensively in the recent literature and older constructions of relational observables where the presence of a heavy background was used to alter the commutator of light operators with the boundary Hamiltonian \cite{Papadodimas:2015xma,Papadodimas:2015jra,Bahiru:2022oas,Bahiru:2023zlc}. With a little translation of notation, the formulas in one set of papers morph into those of the other set!

We defined a trace and an entropy on the gravitationally improved algebra for the time band. The entropy is independent of the characteristics of the observer. More generally, this clarifies that the entropy that has been studied in the context of the gravitational crossed product should be viewed as an effective entropy defined within a little Hilbert space of small excitations about the observer. It should not be conflated with a UV-complete entropy since it is insensitive to the true density of states about the observer in the full theory. 

The algebras for the region $\band$ and $\diam$  as described here are both of type II$_{\infty}$ since the trace of the identity operator diverges . This is because the spectrum of $Q$ is unconstrained both above and below. In contrast, \cite{Jensen:2023yxy} argued that the algebra of a bounded region should be of type II$_1$, which followed from the restriction that the spectrum 
of $Q$ be bounded below.  This assumption was motivated by 
the interpretation of $Q$ as a local observer Hamiltonian whose 
energy was constrained to be positive.  The same restriction
led to the appearance of a type $\ttwo_1$ algebra for the 
static patch of de Sitter space in \cite{Chandrasekaran:2022cip}.  It is of interest to reconcile these constructions and explore whether the imposition of additional physical requirements places constraints on the spectrum of $Q$. 

It is commonly believed that it should be possible to extend the definitions of the gravitational entropy to all orders in perturbation theory in the gravitational coupling. A careful consideration shows that such an extension must surmount various challenges that we have outlined in section \ref{sechigherorder}. These challenges are interesting but not appear insuperable.

\subsection{Irregular regions}

Our paper focused on the algebra of a time band, but there are obvious generalizations of these techniques. For instance, these techniques can be used to define the algebra of an irregularly shaped region in the bulk or the boundary that contains a complete boundary Cauchy slice. In the presence of gravity, the Hamiltonian is one of the observables in this region. The remarks that were applicable to the time band are also applicable to this case: nonperturbatively this algebra comprises all operators in the theory; for simple perturbative excitations about the vacuum, observables in this region suffice to completely fix the state i.e. no information can be hidden from this region. 

But in the presence of a background state with the right properties, we can generalize the constructions of sections \ref{secobserver} and \ref{secalgconst}. We replace polynomials in the boundary operators that appear in \eqref{altcdef} with the set of polynomials in the lowest-order HKLL operators in the irregular region. The definition of the time operator in \eqref{timedef} and \eqref{defq} is expected to remain unchanged. Deforming each element of the algebra using \eqref{hattedopdef} and adding the Hamiltonian should yield the gravitationally corrected algebra corresponding to the region. 

The commutant of the algebra of such an irregular region comprising a boundary Cauchy slice yields the gravitationally corrected algebra of a bulk causal diamond with an irregular boundary.

The remarks above are applicable at leading order in perturbation theory. At subleading order, it is necessary to contend with the challenges outlined in section \ref{sechigherorder}. For an irregular bulk region, there might be additional ambiguities in specifying the algebra.

\subsection{Orthogonal problems}

The techniques used in this paper are widely applicable but, to avoid confusion, we caution the reader that they are {\em not} applicable to some problems in the literature. Indeed, the early discussions of the gravitational crossed product \cite{Witten:2021unn} did not manifest the puzzle discussed in this paper. Those discussions focused on the eternal black hole that has two asymptotic boundaries and studied gravitational corrections to the algebra of one asymptotic boundary. However, the Hamiltonian is an automorphism for such an algebra. Therefore the crossed product emerged without the need of first deforming the nongravitational algebra.

Second, the discussion here is distinct from the question of the consistency of ``islands'' in theories with long-range gravity \cite{Geng:2021hlu}. Islands have commonly been studied in theories where a gravitational theory is coupled to a nongravitational bath and used to obtain a Page curve. (See \cite{Almheiri:2020cfm} and references there.) This bath is not merely a technical tool; it is crucial for defining a factorized Hilbert space and the Page curve that is studied in these setups describes
the transfer of information from one part of the bath to another. Relatedly, the gravitational theory in the presence of a bath is massive \cite{Geng:2020qvw}. 

It has been argued \cite{Raju:2020smc,Raju:2021lwh} that in a theory where gravity is everywhere dynamical, observables outside the black hole always have access to information within the black hole by virtue of the holographic properties that are implicit in gravity. Consequently, while one can obtain a Page curve using various tricks even in these setups, it is not clear whether these curves meaningfully describe information emerging from the black hole. 

One might hope to use the techniques of this paper to circumvent this obstacle. By deforming the operators in the black-hole interior and exterior according to the construction in section \ref{secalgconst}, one might hope to define commuting algebras for the two regions. One might further hope that the entropy for the exterior algebra, as defined in section \ref{secentropy}, would follow a Page curve. 

Such a hope is misguided. The Page curve for black-hole evaporation crucially relies on exponentially suppressed effects of size $\Or[e^{-S}]$. Our construction in section \ref{secalgconst} is a coarse-grained construction that breaks down completely at the nonperturbative level. Moreover, the Page time scales with a power of $N$ which takes us well beyond the regime of this construction. 

It is a conceptual error to apply coarse-grained intuition about local algebras of observables to fine-grained questions about black-hole evaporation. Conflating intuition from these distinct regimes is an invitation for paradox.

We also emphasize that the considerations of this paper do not apply when we study gravitational corrections to the algebra of a nontrivial entanglement wedge that is dual to a subregion on the boundary. The Hamiltonian is never an observable in a nontrivial entanglement wedge and so both bulk and boundary arguments suggest that this algebra is closed even in the presence of gravity without the need for deforming operators. In this case, even nonperturbatively, the algebra is expected to be type $\text{III}_1$ since it is the algebra of a subregion in the boundary quantum field theory. 

An analogue of the dressing that appears here was also studied in the context of semi-infinite time-band algebras in two-sided black holes in \cite{Ali:2024jkx, Faulkner:2024gst}. In \cite{Faulkner:2024gst}, it was noted that the modular dressing causes the algebra of operators in a wedge to the right of the bifurcation surface to fail 
to commute with the left ADM Hamiltonian.  
This complicates the interpretation
of these operators as an algebra associated 
with the right boundary.  It would be helpful to understand whether the ideas presented in the current paper are related and can shed any light on this issue.

\section*{Acknowledgments} We are grateful to Tom Faulkner, Philip H\"ohn, Kyriakos Papadodimas and  Ronak Soni for helpful discussions. We are grateful to participants in the India-South Africa string meeting for helpful discussions;  preliminary results from this work were presented there and we acknowledge the Gravity Theory Trust for its support for this meeting. Research at ICTS-TIFR is supported by the Department of Atomic Energy, Government of India, under Project Identification No. RTI4001.  KJ is supported in part by an NSERC Discovery Grant. 
AJS acknowledges support from the Air Force Office of
Scientific Research under award number FA9550-19-1-036, and from the 
Heising-Simons
Foundation ‘Observational Signatures of Quantum Gravity’ QuRIOS collaboration grant
2021-2817.

\appendix

\section*{Appendix}
\section{Crossed product and mirror operators in the eternal black hole} 
\label{seceternalinterior}

In this appendix, we explain how the construction of mirror operators in the interior of an eternal black hole in \cite{Papadodimas:2015xma,Papadodimas:2015jra} can be framed in the language of a crossed product. The discussion in this Appendix should not be conflated with the discussion of \cite{Witten:2021unn}. The observation in \cite{Witten:2021unn} was that adding the Hamiltonian to the algebra of a single boundary in the thermofield doubled state led to a crossed product algebra. Here, we wish to explain how the crossed product enters in the careful construction of interior operators, which is a different problem.

This construction of \cite{Papadodimas:2015xma,Papadodimas:2015jra} was originally used to demonstrate that state-dependence was necessary even in the context of the  eternal black hole.  The starting observation is as follows. Let $|\tfd \rangle$ be the thermofield doubled state. Then we consider the time-shifted state
\be
\label{timeshifted}
|\tfdT \rangle = e^{-i (H_L + H_R) T/2}  |\tfd \rangle  = e^{-i H_R T}  |\tfd \rangle.
\ee
This time shift is simply a large diffeomorphism acting on the geometry. Therefore it must take a smooth geometry to a smooth geometry and we do not expect it to create a ``firewall'' or another
singularity at the horizon. From the point of view of intrinsic features of the geometry, the states $|\tfdT \rangle$ are on the same footing as $|\tfd \rangle$; the difference between these states is only in how time on the boundary is matched to time in the bulk.

The objective of \cite{Papadodimas:2015xma,Papadodimas:2015jra} was to construct interior operators appropriate for an infalling observer who starts at the right boundary that can be used in the state $|\tfdT \rangle$ for any $T$ and not just in the original state. In particular, these operators should  correctly predict that all such time-shifted states have smooth horizons.
\begin{figure}
\begin{center}
\includegraphics[height=0.3\textheight]{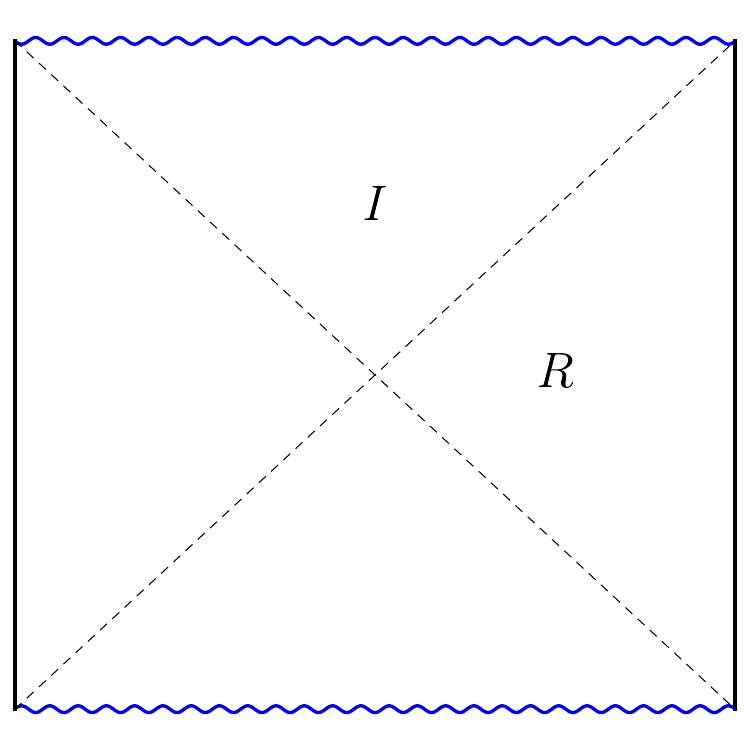}
\caption{\em An eternal black hole. We seek bulk operators that correctly describe the experience of a right-infalling observer in the right exterior (marked R) and the interior (marked I). \label{figeternal}}
\end{center}
\end{figure}

A naive HKLL construction \cite{Hamilton:2005ju} yields operators that are of the form
\be
\label{bulknaive}
\phihkll(t, r, \Omega) = \begin{cases} 
\phihkll^{R}(t, r, \Omega) + \phihkll^L(t, r, \Omega) & \text{if}~(t,r, \Omega)~\text{in~interior}\\
\phihkll^{R}(t, r, \Omega) & \text{if}~(t,r, \Omega)~\text{in~right~exterior} \\
\end{cases}.
\ee
The two regions of interest are marked in Figure \ref{figeternal}. Our notation is slightly schematic: the operators  $\phihkll^{R}(t,r,\Omega)$ and $\phihkll^{L}(t, r, \Omega)$ are obtained by performing an integral transform on operators from the right CFT and the left CFT respectively. The Schwarzschild coordinates $(t,r,\Omega)$ are not smooth everywhere and must be assigned to the different regions in patches. We focus on the right exterior and the future interior geometry since these are the regions accessible to an infalling observer who jumps in from the right boundary.

The integral transforms between boundary and bulk operators are chosen so that the $\phihkll^R$  moves forward in time both inside and outside the horizon under the right Hamiltonian and commutes with the left Hamiltonian
\be
\label{transphir}
e^{i H_R s} \phihkll^{R}(t, r, \Omega) e^{-i H_R s} = \phihkll^{R}(t+s,r, \Omega); \qquad e^{i H_L s} \phihkll^{R}(t, r, \Omega) e^{-i H_L s} = \phihkll^{R}(t,r, \Omega).
\ee
On the other hand, the $\phihkll^L$ operator moves backward in time under the left Hamiltonian and commutes with the right Hamiltonian.
\be
\label{transphil}
e^{i H_R s} \phihkll^{L}(t, r, \Omega) e^{-i H_R s} = \phihkll^{L}(t,r, \Omega); \qquad e^{i H_L s} \phihkll^{L}(t, r, \Omega) e^{-i H_L s} = \phihkll^{L}(t-s,r, \Omega).
\ee

However, \eqref{bulknaive} has a subtle error. This can be seen by studying the correlator
\be
C_{1 2} = \langle \tfdT | \phihkll(t_1, r_1, \Omega_1) \phihkll(t_2, r_2, \Omega_2) | \tfdT \rangle,
\ee
where the point $1$ is in the right exterior just outside the horizon and the point $2$ is in the interior just inside the horizon. When the points are very close, we expect $C_{12} \propto {1 \over s^{k}}$ where $s$ is the geodesic distance between the two points and $k$ depends on the dimension of the operators in the correlator. The short-distance behaviour of this correlator is a diagnostic of whether the geometry is smooth at the horizon. So we expect it to be independent of $T$ since the intrinsic features of the geometry are unaffected by a large diffeomorphism.  Since the two operators transform differently under the time translation that appears in \eqref{timeshifted}, we see that this correlator depends on $T$ and does not have the right short-distance singularity in the time-shifted states.

To correct this, we need operators that transform according to
\be
\label{transformphi}
\begin{split}
&e^{i H_R s} \hat{\phi}(t, r, \Omega) e^{-i H_R s} = \hat{\phi}(t + s,r, \Omega) \\
&e^{i H_L s} \hat{\phi}(t, r, \Omega) e^{-i H_L s} = \hat{\phi}(t ,r, \Omega),
\end{split}
\ee
both inside and outside the horizon. The operators transform under the right Hamiltonian but not the left Hamiltonian because they are dressed to the right boundary. Further discussion can be found in \cite{Papadodimas:2015xma,Papadodimas:2015jra}.

Such operators were constructed in \cite{Papadodimas:2015xma} and then discussed further in \cite{Papadodimas:2015jra}. The construction of that paper, at leading order in $\Or[{1 \over N}]$ can be written using our new notation as follows. The state $ |\tfd \rangle$ plays the role of the state $|\obsst \rangle$. It automatically obeys the conditions that were imposed on this state. These include
\be
E = \langle \tfd | H_R | \tfd \rangle \gg 1; \qquad \text{and} \qquad {\delta E \over E} \ll 1,
\ee
with 
\be
(\delta E)^2 = \langle \tfd | H_R^2 | \tfd \rangle - \langle \tfd | H_R | \tfd \rangle^2 \gg 1.
\ee
Unlike the observer considered in the paper, here we have $E \gg N$. But, as explained above, this poses no obstacle to our construction. The condition \eqref{analcontmatrix} is also obeyed since the thermofield state is thermal.

We proceed as above and introduce a $\Top $ operator that measures the time on the time-shifted states. This $\hat{T}$ operator is shifted forward by both the left and the right Hamiltonian
\be
\label{transT}
e^{i H_R s} \hat{T} e^{-i H_R s} = e^{i H_L s} \hat{T} e^{-i H_L s} = \hat{T} + s.
\ee

Then we switch the commutator of the left operators in the interior by twirling them with an appropriate unitary
\be
\label{correctinterior}
\widehat{\phi}(t, r, \Omega) = \begin{cases} 
\phihkll^{R}(t, r, \Omega) + e^{i (H_R - H_L) \Top } \phihkll^L(t,r, \Omega) e^{-i (H_R - H_L) \Top }  & \text{if} ~(t,r,\Omega)~\text{in~interior}\\
\phihkll^{R}(t, r, \Omega) & \text{if}~(t,r,\Omega)~\text{in~right~exterior} \\
\end{cases}.
\ee
The operators \eqref{correctinterior} improve \eqref{bulknaive} and provide the correct description of operators in the interior of the eternal black hole.  It is not hard to check that the relations \eqref{transformphi} are satisfied using the commutators \eqref{transphir}, \eqref{transphil} and \eqref{transT}. 

By expanding the unitaries used to twirl the operator in terms of projectors onto the time-shifted little Hilbert spaces, we see that this precisely matches the construction of \cite{Papadodimas:2015xma, Papadodimas:2015jra} at leading order in $\Or[{1 \over N}]$.  

The twirled left operators that appear above are what are called ``mirror operators'' in the literature on bulk reconstruction. We see that the mirror operators belong to a crossed product algebra. This is a modular crossed product since $H_R - H_L$ is the modular Hamiltonian in the state. It is amusing to note that the crossed product had appeared secretly in this old construction. 

In conclusion, we should mention that the discussion of \cite{Papadodimas:2015xma, Papadodimas:2015jra} went beyond the leading order approximation. Those papers discussed the exponentially small terms in the overlap between $\langle \tfdT | \tfd \rangle$ (which was later related to the spectral form factor in \cite{Cotler:2016fpe} ). These overlaps were estimated and it was argued that if one seeks operators that are valid for an exponentially large range of $T$ then such operators must be state dependent. In this Appendix, consistent with the spirit of the main text of this paper, we have not concerned ourselves with these nonperturbative effects.

\bibliographystyle{JHEPthesis}

\bibliography{references}

\end{document}